\pdfoutput=1

\documentclass{article}
\usepackage{arxiv}

\usepackage{subcaption}

\usepackage{lineno,hyperref}
\usepackage{amsthm}
\usepackage{amsmath,amssymb}
\usepackage{subcaption}
\usepackage{stmaryrd}
\usepackage{bm}
\usepackage{float}
\usepackage{adjustbox}
\usepackage{tabularx}
\usepackage{multirow}
\usepackage{amsfonts}
\usepackage{mathtools}

\usepackage{amsfonts}

\DeclareMathAlphabet\mathbfcal{OMS}{cmsy}{b}{n}
\usepackage{diagbox}
\newtheorem{prop}{Proposition}
\newtheorem{defi}{Definition}
\newtheorem{theorem}{Theorem}
\newtheorem{remark}{Remark}
\usepackage{lineno}

\usepackage{cleveref}
\title{A Quasi-Newton method for physically-admissible simulation of Poiseuille flow under fracture propagation.}

\author{
  Guotong Ren\\
  Department of Petroleum Engineering\\
  University of Tulsa\\
  Tulsa, OK 740104 \\
  \texttt{guotong-ren@utulsa.edu} \\
   \And
 Rami M.~Younis \\
  Department of Petroleum Engineering\\
  University of Tulsa\\
  Tulsa, OK 74104 \\
  \texttt{rami-younis@utulsa.edu} \\
}
\begin{document}
\maketitle
\begin{abstract}{Coupled hydro-mechanical processes are of great importance to numerous engineering systems, e.g., hydraulic fracturing, geothermal energy, and carbon sequestration. Fluid flow in fractures is modeled after a Poiseuille law that relates the conductivity to the aperture by a cubic relation. Newton's method is commonly employed to solve the resulting discrete, nonlinear algebraic systems. It is demonstrated, however, that Newton's method will likely converge to nonphysical numerical solutions, resulting in estimates with a negative fracture aperture. A Quasi-Newton approach is developed to ensure global convergence to the physical solution. A fixed-point stability analysis demonstrates that both physical and nonphysical solutions are stable for Newton's method, whereas only physical solutions are stable for the proposed Quasi-Newton method. Additionally, it is also demonstrated that the Quasi-Newton method offers a contraction mapping along the iteration path. Numerical examples of fluid-driven fracture propagation demonstrate that the proposed solution method results in robust and computationally efficient performance.}
\end{abstract}


\maketitle

\section{Introduction}
\begin{figure}[ht]
	\centering
	\begin{subfigure}[t]{0.49\textwidth}
		\centering
		\includegraphics[width=0.6\linewidth]{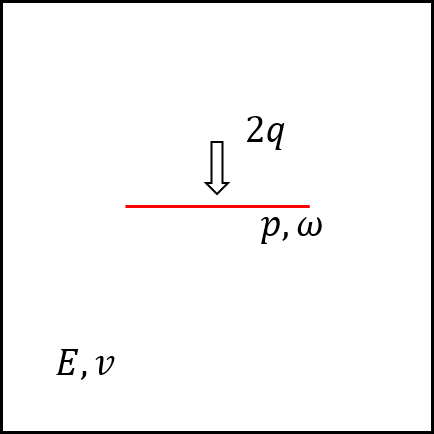}
		\caption{}
		\label{fig:singlefrac}
	\end{subfigure}
	\begin{subfigure}[t]{0.49\textwidth}
		\centering
		\includegraphics[width=1\linewidth]{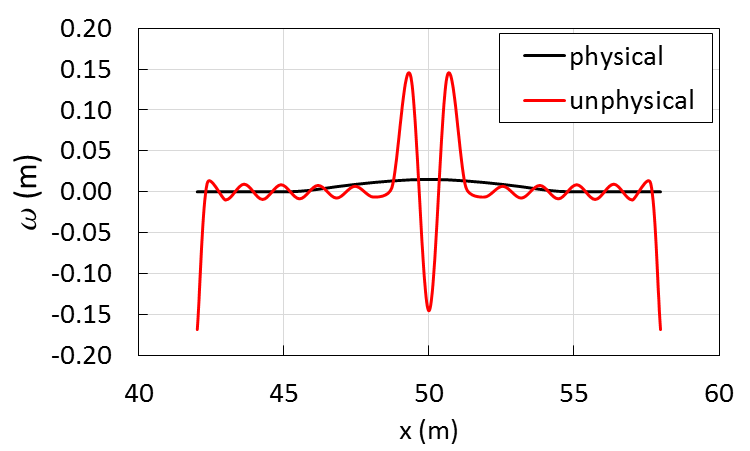}
		\caption{}
		\label{fig:NegAper}
	\end{subfigure}
	\caption{Example of multiple solutions:~(a) case schematic;~(b) aperture profile. The results are generated using coupled XFEM-FVM~\cite{ren2021integrated}}
	\label{fig:discretization}
\end{figure}

In hydro-mechanical processes, the interplay between rock deformation and fluid pressure in fractures can dictate first-order effects in several engineering systems such as hydraulic fracturing~\cite{bavzant2014fracking,liu2019history}, geothermal utilization~\cite{breede2013systematic}, and  $\text{CO}_2$ sequestration~\cite{edwards2015model,xu2021revisiting}. In the limits of linearity, and assuming a constant positive-definite permeability tensor in fracture, existence and uniqueness of solution to the governing continuity equations has been established ~\cite{girault2015lubrication,girault2016convergence}. The physically-accepted limit of Poiseuille flow introduces a nonlinear relation for fracture conductivity as a cubic function of local aperture. While formal conditions for the existence and uniqueness of solutions are well-established, numerical and semi-analytical  approximations are widely applied. Newton's method is commonly employed to solve resulting algebraic equations whether it be in the contexts of fully- or sequentially-coupled solution processes. Numerical convergence difficulties can arise, as is generally the case with black-box nonlinear solution processes. More notably however, it is observed that the Newton iteration can converge to nonphysical solutions (i.e. aperture is negative). This is a frequently occurring issue and numerical evidence is easily obtained. Consider for example, a simple two-dimensional model with a fracture located at the center of a domain, and where fluid is injected into the mid-point of the fracture. Initially, the scaled fluid pressure and displacement fields are zero. Two alternate aperture profiles along the fracture are presented in~\cref{fig:NegAper}; while both solutions satisfy the same nonlinear convergence criteria for the same model, one profile is oscillatory and includes negative aperture, whereas the other does not. Negative fracture (aperture) conductivity violates the positive-definite tensor requirement of \cite{girault2015lubrication,girault2016convergence},  and permits flow from low fluid pressure to high fluid pressure. This simple example may support the hypothesis that the hydro-mechanical formulation possesses multiple solutions and that the nonphysical one can be obtained by application of Newton's method. We omit the specifics of the numerical approximation and physical parameters employed in this illustrative example; in a subsequent section, it is demonstrated that the nonphysical solution exists regardless of the discretization employed. 

While significant to robust numerical simulation, the literature on the aforementioned issues is scarce. In this article, our focus is on linear elastic fracture propagation problems without fluid lag; i.e., fluid is assumed to occupy the entire fracture space at each instance in which the fracture tip reaches a failure state. Two classes of numerical fracture propagation simulation algorithm are: (1) Given a fixed tip advancement length step, determine the time increment such that tip is in an equilibrium failure state, and the solution satisfies the continuity equations (e.g., \cite{gupta2018coupled,liu2020modeling,ren2021integrated}); and (2) given a time step, determine the advancement length step for each tip such that the solution fulfills equilibrium failure (e.g., \cite{gordeliy2013implicit,hunsweck2013finite}). In both classes of algorithm, the nonphysical solution may be encountered. \cite{gupta2018coupled, ren2021integrated} propose initialization schemes for unknowns at every time that new fracture segments are introduced based on surrogate models. Nevertheless, while these initial guesses may improve convergence, it is observed that convergence to a nonphysical solution remains likely to occur in viscosity-dominated fracture propagation. \cite{girault2015lubrication} propose a safeguarding strategy by applying a relaxation to $\omega$ within the iterative coupled scheme. As $\omega < 0$ is detected during nonlinear iterations, the $\omega$ iterate is manually reassigned to a non-negative value. Although such an approach ensures a positive fracture conductivity should the method converge, an open question pertains to the effects on convergence rate and reliability. 

A Quasi-Newton method is proposed to address global convergence to positive aperture solutions. The method proposes a Jacobian matrix modification that is inspired by that used in nonlinear two-point-flux finite volume methods (e.g. \cite{lipnikov2009interpolation,terekhov2017cell}). For example in~\cite{lipnikov2009interpolation,terekhov2017cell}, in order to maintain a monotone Jacobian matrix, derivatives of the transmissibility with respect to  pressure are neglected. As a result, the converged pressure solutions are assured to be non-negative. For the fracture propagation problem and coupled hydro-mechanical problems at hand, monotonicity of the linearized operator is not guaranteed. For instance, the physical pressure solution may take on negative values near the tip region in the viscosity-dominated fracture propagation regime (fluid lag zones for example). Subsequently, the proposed approach will demonstrate that: (1) the fixed-point stability for the Newton's and proposed Quasi-Newton methods are different; (2) the proposed Quasi-Newton method provides a contraction mapping property with Lipschitz constant between $0$ and $1$ on a designed iteration path; and (3) a non-negative aperture $\omega >= 0$ is ensured at every nonlinear iteration.
   
Next, the mathematical formulation of the problem is detailed in~\cref{sec:PS}. Then, in~\cref{sec:NS}, the Quasi-Newton method is proposed and analyzed under the setting of an infinite domain where the mechanics equation can be formulated using a closed Green's function between $\omega$ and pressure $p$. In~\cref{sec:XFEM}, the development is extended to more general fracture configurations. Specifically, we extend the algorithm to discretization using a coupled extended-finite-element method (XFEM) and finite volume method (FVM) for mechanics and flow, respectively~\cite{ren2021integrated}. 

\section{Problem statement}\label{sec:PS}
Consider a spatial domain $\Omega\subset \mathbb{R}^2$ with external boundary $\Gamma$ and its associated outward-oriented unit-normal $\bm{n}_{\Gamma}$. Dirichlet and Neumann boundaries for mechanics are $\Gamma_{u}$ and $\Gamma_{t}$. The boundary segments are disjoint ($ \Gamma_{u} \bigcap \Gamma_{t} = \varnothing$), and $\Gamma_{u} \cup \Gamma_{t} = \Gamma$.  

A one-dimensional lower representation of the fracture, $\mathcal{C}$, is adopted. Fracture aperture, $\omega$, is defined as,
\begin{equation}
\omega = \llbracket \boldsymbol{u} \rrbracket \cdot \boldsymbol{n}_{c}.
\end{equation}
where $\llbracket \boldsymbol{u} \rrbracket$ is the jump of the displacement over the fracture, $\bm{n}_{c}$ is the fracture unit-normal vector. 

A single-phase incompressible fluid occupies fracture, and linear elastic mechanical deformation model is assumed. The continuity equation in the fracture is
\begin{equation}\label{govM}
\frac{\partial\omega}{\partial t} - \nabla_c \cdot \left(\frac{\omega^{3}}{12\mu}\nabla_c p\right) = 0 \quad \text{on} \quad \mathcal{C},
\end{equation}
where $\nabla_c$ is the gradient operator defined on the fracture path. $p$ is the fluid pressure in the fracture and $\mu$ is the fluid viscosity. Poiseuille's law posits that the fracture conductivity is $\frac{\omega^{3}}{12}$. The boundary condition at the inlet of the fracture is 
\begin{align}
-\frac{\omega^{3}}{12\mu}\frac{\partial p}{ \partial x} = Q, 
\end{align}
where $Q$ is the fluid injection rate, and at the tip
\begin{align}
-\frac{\omega^{3}}{12\mu}\frac{\partial p}{ \partial x} = 0.
\end{align}
The quasi-static momentum equation reads
\begin{equation}
\label{GeomechanicsSF}
\nabla \cdot \bm{\sigma}=\boldsymbol{0}, \quad \text{on}\quad \Omega,
\end{equation}
where $\bm{\sigma}$ is the second order tensor. On the external boundary $\Gamma$, Neumann (force) and Dirichlet (displacement) conditions are
\begin{subequations}
	\begin{align}
	\bm{\sigma}\cdot\bm{n}_{\Gamma}= \bm{t}  \quad &\text{on}\quad\Gamma_{u},\\
	\bm{u} = \bm{\hat{u}}   \quad  &\text{on}\quad \Gamma_{t},
	\end{align}
\end{subequations}
while on immersed fracture boundaries, $p$ is imposed onto the oriented surfaces of the fracture:
\begin{subequations}
    \label{innerBC}
	\begin{align}
	\bm{\sigma}\cdot \bm{n_{c}} =p\boldsymbol{I}\cdot\bm{n_{c}} \quad &\text{on}\quad\mathcal{C}.
	\end{align}
\end{subequations} 
The stress $\bm{\sigma}$ is modeled using linear elastic theory:
\begin{equation}\label{stress_strain}
\bm{\bm{\sigma}} = \frac{E\nu}{(1+\nu)(1-2\nu)} (\nabla \cdot \bm{u}) \bm{I} + \frac{E}{1+\nu}\bm{\varepsilon},
\end{equation}
where $\boldsymbol{I}$ is the identity matrix, $E$ is Young's modulus, $\nu$ is Poisson's ratio and the strain $\boldsymbol{ \varepsilon}$ is a second order tensor. Under infinitesimal deformation, the strain tensor is a function of displacement as 
\begin{equation}\label{strain}
\bm{\varepsilon} =   \frac{1}{2}(\nabla^{\text{T}} \bm{u} + \nabla \bm{u}).
\end{equation}

\subsection{Numerical Discretization}
Two discretization schemes will be employed in computational examples which are referred to as \textit{DS1} and \textit{DS2}. DS1 considers a single fracture in an infinite domain while DS2 deals with rather general fracture and domain geometry. The simulation test performed only considers the static fracture. The propagation scenario will be investigated in the next section. Both schemes are briefly introduced.
\subsubsection{DS1}
A single fracture is modeled in an infinite domain as shown in \cref{fig:singlefrac}. Due to symmetry, only half of the domain is modeled. According to \cite{spence1985self}, the aperture field $\omega$ is obtained by solving~\cref{GeomechanicsSF,innerBC,stress_strain,strain}, and can be written as an explicit function of $p$ by the Green's function $\mathcal{G}$,
\begin{equation}\label{greenfunc}
\begin{split}
    \omega(x) &= -\frac{2(1-\nu^2)}{\pi E}\int_{0}^{a} \mathcal{G}(s; x)p(s)ds\\
    &=  -\frac{2(1-\nu^2)}{\pi E}\int_{0}^{a} \ln{\left|\frac{(a^2 - x^2)^{\frac{1}{2}} - (a^2 - s^2)^{\frac{1}{2}}}{(a^2 - x^2)^{\frac{1}{2}} + (a^2 - s^2)^{\frac{1}{2}}} \right|}p(s)ds
\end{split}
\end{equation}
where $a$ is the fracture half length. Consequently, an integro-differential equation system, \cref{govM,greenfunc}, is formulated. Next, a discrete approximation of this formulation is described.

The domain $\mathcal{C}$ is approximated with $n_c$ equally-spaced grid cells using length equal to $\Delta x = \frac{a}{n_c}$. $\boldsymbol{\omega} = \{\omega_i, i = 1,...,n_c\}$ and $\boldsymbol{p} = \{p_i, i = 1,...,n_c\}$, both of which are co-located at the center of $n_c$ cells. A Gaussian quadrature integration rule is used to evaluate $\omega_i$,
\begin{equation}\label{discretizedM}
\begin{split}
    \omega_i\big(x_i = \frac{2i - 1}{2}\Delta x\big)& = -\frac{2(1-\nu^2)}{\pi E}\sum_{j = 1}^{n_c}\int_{(j-1)\Delta x}^{j \Delta x} \mathcal{G}(s;x_i)p_jds \\
    &= -\frac{2(1-\nu^2)}{\pi E} \sum_{j = 1}^{n_c}\sum_{k = 1}^{n_{g}} \mathcal{G}(s^{*}_k;x_i)\mathcal{W}_kp_j 
\end{split}
\end{equation}
where $n_g$ is the number of quadrature points; $s_k^*$ is the quadrature point; $\mathcal{W}_{k}$ is the weight at the point $s_j^*$. \cref{discretizedM} can be simplified using a matrix-vector notation,
\begin{equation}\label{matrixomega}
    \boldsymbol{\omega} = \boldsymbol{A}\boldsymbol{p}
\end{equation}
where each item in full matrix $\boldsymbol{A}$ stores the numerical integration result of each cell. Since for all $ s \text{ and } x \subseteq (0,a)$, $\mathcal{G} (s;x) < 0$, \text{then } $A_{ij} > 0$.

Pertaining to fluid flow, a first-order backward Euler temporal discretization, and a second-order central-difference spatial discretization are applied. The discretized form reads,
\begin{equation}\label{discretizedflow}
    (\omega^{n+1}_{i} - \omega_{i}^{n}) - T\sum_{\pm} \left(\omega^{n+1}_{i\pm \frac{1}{2}}\right)^3(p^{n+1}_{i\pm 1} - p^{n+1}_{i}) = 0    
\end{equation}
where superscripts $n$ or $n+1$ indicate time level, and $\Delta t = t^{n+1} - t^{n}$; $T = \frac{\Delta t}{12\mu\Delta x^2}$ is the static transmissibility while $\omega$ at the interface is evaluated by the arithmetic average of its neighboring cells', i.e. $\omega_{i\pm \frac{1}{2}} = \frac{\omega_{i} + \omega_{i\pm 1}}{2}$. Combining \cref{matrixomega,discretizedflow}, a matrix-vector form of the nonlinear system is formulated
\begin{equation}\label{nonlinearSystem}
    \big(\boldsymbol{A} + \boldsymbol{F}(\boldsymbol{p}^{n+1})\big)\boldsymbol{p}^{n+1} = \boldsymbol{q} + \boldsymbol{\omega}^{n}
\end{equation}
where $\boldsymbol{q} = (\frac{\Delta t}{\Delta x} Q, 0,...,0)^T$ is a $n_c$ by $1$ vector, $\boldsymbol{F}$ represents the flux term and is a tri-diagonal sparse matrix under a 2D setting. The detailed expression of $\boldsymbol{F}$ is given in \ref{appendix2}.

\subsubsection{DS2}
In general scenarios with multiple fractures, as well as complex-fracture geometries, the original problem is approximated using a mixed discretization;  coupled extended FEM and embedded FVM for mechanics deformation and fluid flow, respectively. The numerical details of the method is referred to the \cref{Discretization}.

\section{Proposed Quasi-Newton Solver}\label{sec:NS}
Newton's method is typically applied to solve nonlinear algebraic systems that arise from discretization. It is observed that these methods may converge to nonphysical solutions. In the following subsections, this is analyzed theoretically and empirically in the contexts of models DS1 and DS2 introduced above. Furthermore, a Quasi-Newton approach is proposed, and demonstrated to provide robust global convergence to the physical solution.

\subsection{Analysis and development using model DS1}\label{sec:quasi-newton}
The nonlinear system~\cref{nonlinearSystem} is to be solved, where the residual vector $\boldsymbol{R}$ and Jacobian matrix $\boldsymbol{J}$ in the context of DS1 become,
\begin{subequations}\label{eq:quasi-newton}
\begin{align}
    &R^{n+1, v+1}_{i} = (\omega^{n+1,v}_{i} - \omega_{i}^{n}) - T\sum_{\pm} \left(\omega^{n+1,v}_{i\pm \frac{1}{2}}\right)^3(p^{n+1,v}_{i\pm 1} - p^{n+1,v}_{i}) - q_i  \label{eq:residual}  \\
    \begin{split}
     &J^{n+1, v+1}_{ij} = \frac{\partial \omega^{n+1,v}_i}{\partial p^{n+1,v}_j} - T\sum_{\pm} \left(\omega^{n+1,v}_{i\pm \frac{1}{2}}\right)^3\frac{\partial(p^{n+1,v}_{i\pm 1} - p^{n+1,v}_{i})}{\partial p_j^{n+1,v}} \\
     &- T\sum_{\pm} \frac{\partial \left(\omega^{n+1,v}_{i\pm \frac{1}{2}}\right)^3}{\partial p^{n+1,v}_j}(p^{n+1,v}_{i\pm 1} - p^{n+1,v}_{i}) \label{eq:Jacobian}\end{split}\\
     &\boldsymbol{J}^{n+1, v+1} \delta \boldsymbol{p}^{n+1, v+1} = -\boldsymbol{R}^{n+1, v+1} \label{standardNewton}
\end{align}
\end{subequations}
where superscripts $v$ or $v+1$ represent the iteration level. The first term on the right-hand side (RHS) of \cref{eq:Jacobian}, $\frac{\partial \omega^{n+1,v}_i}{\partial p^{n+1,v}_j}$, is equal to $A_{ij}$; the second term on the RHS of \cref{eq:Jacobian} forms a diagonally dominant sparse matrix whose diagonal terms are positive and off-diagonal terms are negative. The last term on the RHS of \cref{eq:Jacobian} produces a full matrix with elements consisting of derivatives of $\omega^3$ with respect to $p$. The subsequent numerical study demonstrates that the presence of this term can lead to convergence of Newton's method to a nonphysical solution. In the proposed Quasi-Newton approach, the third term is neglected in the Jacobian calculation. Note that in \cite{lipnikov2007monotone,terekhov2017cell}, a similar strategy is adopted but for the purpose of guaranteeing a monotone linear matrix. The QN method now reads,
\begin{subequations}
\begin{align}
     &\tilde{J}^{n+1, v+1}_{ij} = \frac{\partial \omega^{n+1,v}_i}{\partial p^{n+1,v}_j} - T\sum_{\pm} \left(\omega^{n+1,v}_{i\pm \frac{1}{2}}\right)^3\frac{\partial(p^{n+1,v}_{i\pm 1} - p^{n+1,v}_{i})}{\partial p_j^{n+1,v}} \\
    &\tilde{\boldsymbol{J}}^{n+1, v+1}\boldsymbol{p}^{n+1,v+1} =  \big(\boldsymbol{A} + \boldsymbol{F}(\boldsymbol{p}^{n+1, v})\big)\boldsymbol{p}^{n+1,v+1} = \boldsymbol{q} + \boldsymbol{\omega}^{n} \label{quasinewton}
\end{align}
\end{subequations}
From \cref{quasinewton}, a mapping $K: X \rightarrow X, X \subset \mathbb{R}^{n_c}$ is defined as
\begin{equation}\label{quasi-newton_mapping}
   K = \big(\boldsymbol{A} + \boldsymbol{F}(\boldsymbol{p})\big)^{-1}(\boldsymbol{q} + \boldsymbol{\omega}^{n}) 
\end{equation}
Consequently, our Quasi-Newton \cref{quasinewton} searches for fixed points over set $X$. 
The following proposition proposes a property on set $X$, 
\begin{prop}\label{prop1}
$\forall \boldsymbol{p} \in X$, mapping $K : X\rightarrow X$, and $X \subset \mathbb{R}^{n_c}$ is a vector space that satisfies $\sum_i(\boldsymbol{A}\boldsymbol{p}^*)_i = \sum_i(\boldsymbol{q} + \boldsymbol{\omega}^{n})_i, \text{ where } \boldsymbol{p}^* = K(\boldsymbol{p})$.
\end{prop}
Proposition \ref{prop1} states that the total mass balance is satisfied during each iteration. The proof of the Proposition \ref{prop1} is in Appendix \ref{appendix1}. 
\subsubsection{Analysis of fixed-point stability}
\begin{figure}[htb]
    \begin{subfigure}{0.49\textwidth}
        \centering
		\includegraphics[width=1\linewidth]{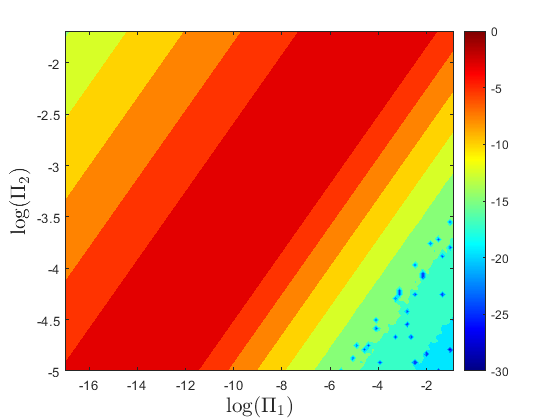}
		\caption{}
    \end{subfigure}
    \begin{subfigure}{0.49\textwidth}
        \centering
		\includegraphics[width=1\linewidth]{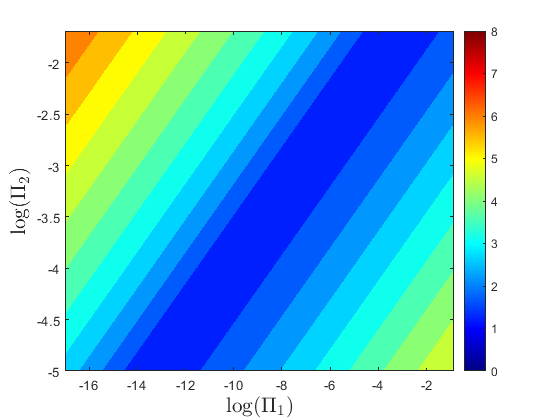}
		\caption{}
    \end{subfigure}
    \caption{$\log(\rho(K'))$ of Quasi-Newton method v.s $\left(\log(\Pi_1), \log(\Pi_2)\right)$. (a) physical solution; (b) nonphysical solution. }
    \label{fig:qn_spectral}
\end{figure}

\begin{figure}
    \centering
    \includegraphics[width=0.55\linewidth]{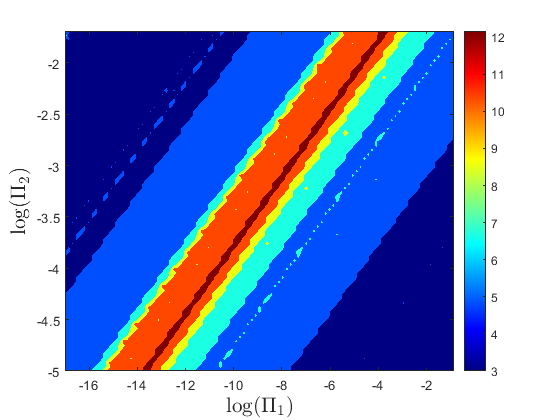}
    \caption{nonlinear iterations of each case using the slightly perturbed nonphysical solution as the initial guess}
    \label{fig:stability_niter}
\end{figure}

There may exist multiple fixed-points in $X$. The stability of such fixed-points for nonlinear mappings $K$ corresponding to both Newton's and Quasi-Newton methods is analyzed. The stability of fixed points is defined by Definition \ref{stability}.
\begin{defi}\label{stability}Fixed-points can be classified as,
\begin{itemize}
    \item The fixed point $\boldsymbol{p}_0$ is stable if there exists an open set $U\subset X$ containing $\boldsymbol{p}_0$ such that $\|K(\boldsymbol {p}) - \boldsymbol{p}_0 \| \leq \| \boldsymbol{p} - \boldsymbol{p}_0\|$ for all $\boldsymbol{p} \in U$
    \item The fixed point $\boldsymbol{p}_0$ is unstable if there exists an open set $U\subset X$ containing $\boldsymbol{p}_0$ such that $\|K(\boldsymbol {p}) - \boldsymbol{p}_0 \| \geq \| \boldsymbol{p} - \boldsymbol{p}_0\|$ for all $\boldsymbol{p} \in U$
\end{itemize}
\end{defi}
A well-studied result in discrete dynamics characterizes the stability of fixed-points for multi-variable nonlinear operators.
\begin{theorem}(see \cite{campos2017stability})
Suppose map $K: X\rightarrow X$ is differentiable at a fixed point $\boldsymbol{p}_0$ and then let $\lambda_1, \lambda_2,...,\lambda_n$ be the eigenvalues of the Jacobian matrix $K'$ evaluated at $\boldsymbol{p}_0$.
\begin{itemize}
    \item if all the eigenvalues $\lambda_j$ have $|\lambda_j|< 1$, then $\boldsymbol{p}_{0}$ is stable or attracting.
    \item if one eigenvalue $\lambda_{j0}$ has $|\lambda_{j0}|> 1$, $\boldsymbol{p}_{0}$ is unstable, which can be either saddle or repelling.
    \item if all the eigenvalues $\lambda_j$ have $|\lambda_j|> 1$, then $\boldsymbol{p}_{0}$ is repelling.
\end{itemize}
\end{theorem}
An empirical analysis of the stability of the fixed-points is conducted. The Buckingham $\pi$ theorem is applied in a dimensional analysis to yield the following dimensionless groups,
\begin{equation}
    \Pi_1 = \frac{\mu}{Et}, \quad \Pi_2 = \frac{q t}{a^2}, \quad \Pi_3 = \frac{p}{E}
\end{equation}
Based on the Buckingham $\pi$ theorem, the solution $\Pi_3$ only depends on the magnitude of $\Pi_1$ and $\Pi_2$. The domain setup is shown in \cref{fig:singlefrac}. A single fracture is in the center of an infinite domain. Initially, the fracture is void of fluid, i.e. $\boldsymbol{\omega}^n = 0$. The simulation runs for one time step $\Delta t = t$. The fracture is represented using four grid cells. The input parameter $\Pi_1$ is varied across simulations from $1e-17$ to $1.5e-1$ and $\Pi_2$ from $1e-5$ to $2e-2$. As a result, there are 8000 simulation cases in total. The physical and nonphysical solutions are generated for each one of 8000 simulation cases. Subsequently, then spectral radius of $K'$, $\rho(K')$ is evaluated for each case at the physical and nonphysical solutions. These results are illustrated in \cref{fig:qn_spectral}. All nonphysical solutions are unstable fixed-points of the Quasi-Newton mapping, while all physical solutions are stable. In other words, if the Quasi-Newton method converges, it will not do so to a nonphysical solution since any small perturbation will push the next iterate away from the unstable fixed point. To demonstrate this, we apply the nonphysical solution perturbed by $0.01\%$ as an initial guess for the Quasi-Newton method. All 8000 cases converge to the physical solution, with a maximum of 12 nonlinear iterations. The number of nonlinear iterations required to convergence over the entire test-set appears in~\cref{fig:stability_niter}. 

\begin{figure}[htb]
    \begin{subfigure}{0.49\textwidth}
        \centering
		\includegraphics[width=1\linewidth]{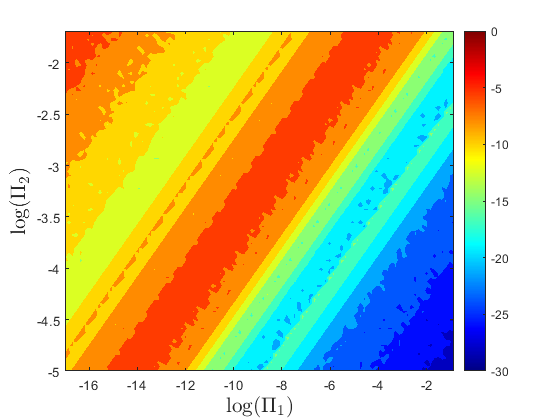}
		\caption{}
    \end{subfigure}
    \begin{subfigure}{0.49\textwidth}
        \centering
		\includegraphics[width=1\linewidth]{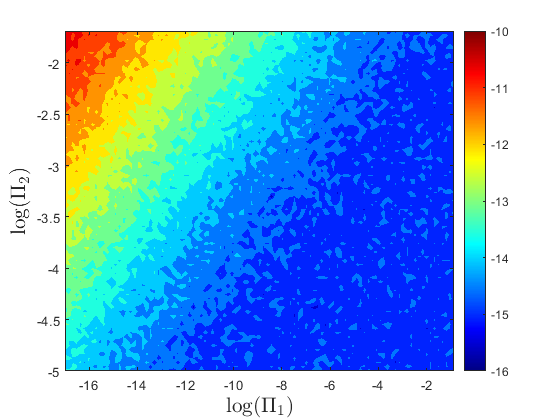}
		\caption{}
    \end{subfigure}
    \caption{$\log(\rho(K'))$ of Newton's method v.s $\left(\log(\Pi_1), \log(\Pi_2)\right)$. (a) physical solution; (b) nonphysical solution. }
    \label{fig:n_spectral}
\end{figure}

Similarly, the stability of fixed-points for the Newton operator is also investigated. A mapping $K$ for Newton's method can be defined as,
\begin{equation}
K(\boldsymbol p) = \boldsymbol{p} - \boldsymbol{J}^{-1}(\boldsymbol p)\boldsymbol{R}(\boldsymbol p)
\end{equation}
The computed spectral radius $\rho(K')$ at physical and nonphysical solutions over the investigation space are presented in \cref{fig:n_spectral}. Clearly, both physical and nonphysical solutions are stable fixed points in $X$. This explains why Newton's method may converge to the negative aperture solution when applied to any of these problems.
\subsubsection{contraction mapping}
\begin{defi}
Let $(D, d)$ define a metric space. Mapping $G : D \rightarrow D$ is a contraction, if there exists a constant $c$, with $0< c < 1$, such that 
\begin{equation}\label{eq:contraction}
    d\big(G(y_1) - G(y_2)\big) < c d(y_1, y_2)
\end{equation}
for all $y_1, y_2 \in D$. Note that $c$ in \cref{eq:contraction} is the Lipschitz constant.
\end{defi}
 Next, it is demonstrated that a mapping $G$ for $\omega$ preserves the contraction mapping property on a certain iteration path. The RHS of \cref{quasi-newton_mapping} is left multiplied by matrix $A$, to produce $G$ as,
 \begin{equation}\label{eq:aperture_mapping}
     G =  \boldsymbol{A}(\boldsymbol A+\boldsymbol F(\boldsymbol \omega))^{-1}(\boldsymbol{q} + \boldsymbol{\omega}^{n})
 \end{equation}
Firstly, we define a set $D$ that only contains physical solutions.
\begin{defi}
Set $D$ is defined such that dimensionless aperture $\frac{\boldsymbol{\omega}}{\sqrt{Qt}} \geq \varepsilon_0$. Here $\varepsilon_0$ is a fixed tolerance whose absolute value, $\mid \varepsilon_0 \mid$, is close to $0$.
\end{defi}
In order to show there exists a $0<c<1$ such that mapping $G: D \rightarrow D$ is a contraction, we design an iteration path for mapping $G$
\begin{equation}\label{eq:contraction_g}
    G = \boldsymbol A \big(\boldsymbol{A} + \boldsymbol{F}(\boldsymbol{\omega}^{n+1, v})\big)^{-1}(\boldsymbol{q} + \boldsymbol{\omega}^{n})
\end{equation}
where $v$ stands for applying $G$ for $v$ iterations, and $\omega^{n+1, v}$ is 
\begin{equation}\label{eq:contraction_g_path}
\omega^{n+1, v} = \left\{\begin{matrix}
&\omega^{n} & v=1\\ 
&\omega^{n+1, v-1} & v\geq 1
\end{matrix}\right.
\end{equation}
where in the first iteration, the previous time-step solution is applied as the initial guess.

There are two things that need to be demonstrated for a contraction mapping: (1) if $\frac{\boldsymbol{\omega}}{\sqrt{Qt}} \in D$, $\frac{K(\boldsymbol{\omega})}{\sqrt{Qt}} \in  D$; (2) there exists $0<c<1$ on the designed iteration path, \cref{eq:contraction_g} and \cref{eq:contraction_g_path}. To present the value of $c$ during the numerical simulation, the following calculation is adopted
\begin{equation}
    c = \frac{d (G^v (\omega^v), G^v(\omega^{v-1})) }{d(\omega^v, \omega^{v-1})} = \frac{d(\omega^{v+1} ,\omega^{v}) }{d(\omega^v , \omega^{v-1})}
\end{equation}
where operator $d$ is chosen as $L_2$ norm in this manuscript.

\begin{figure}[htb]
\centering
    \begin{subfigure}{0.49\textwidth}
        \centering
		\includegraphics[width=1\linewidth]{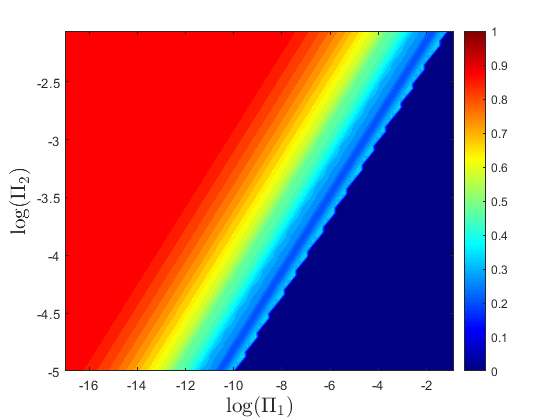}
		\caption{}
		\label{fig:lipschitz}
    \end{subfigure}
    \begin{subfigure}{0.49\textwidth}
        \centering
		\includegraphics[width=1\linewidth]{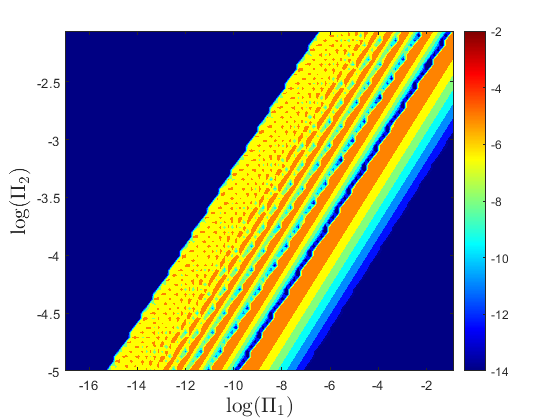}
		\caption{}
		\label{fig:omega}
    \end{subfigure}
    \begin{subfigure}{0.49\textwidth}
        \centering
		\includegraphics[width=1\linewidth]{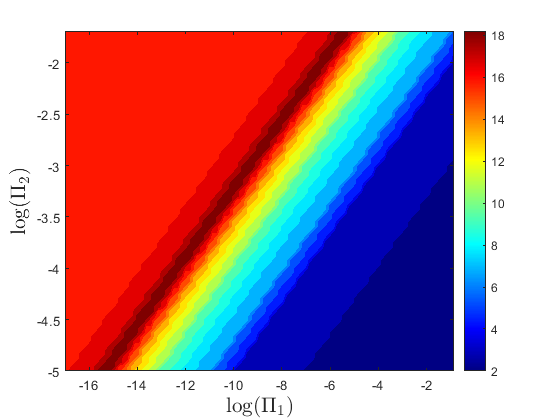}
		\caption{}
		\label{fig:niter}
    \end{subfigure}
    \caption{Quasi-Newton Contraction Verification: (a) the maximum $c$ during nonlinear iterations; (b) the minimum dimensionless $\frac{\omega}{\sqrt{qt}}$ during nonlinear iterations and  $\log(-\frac{\omega}{\sqrt{Qt}})$ is presented; (c) nonlinear iterations for each case.}
    \label{fig:contraction}
\end{figure}

\begin{figure}[ht]
	\centering
	\begin{subfigure}[t]{0.49\textwidth}
		\centering
		\includegraphics[width=1\linewidth]{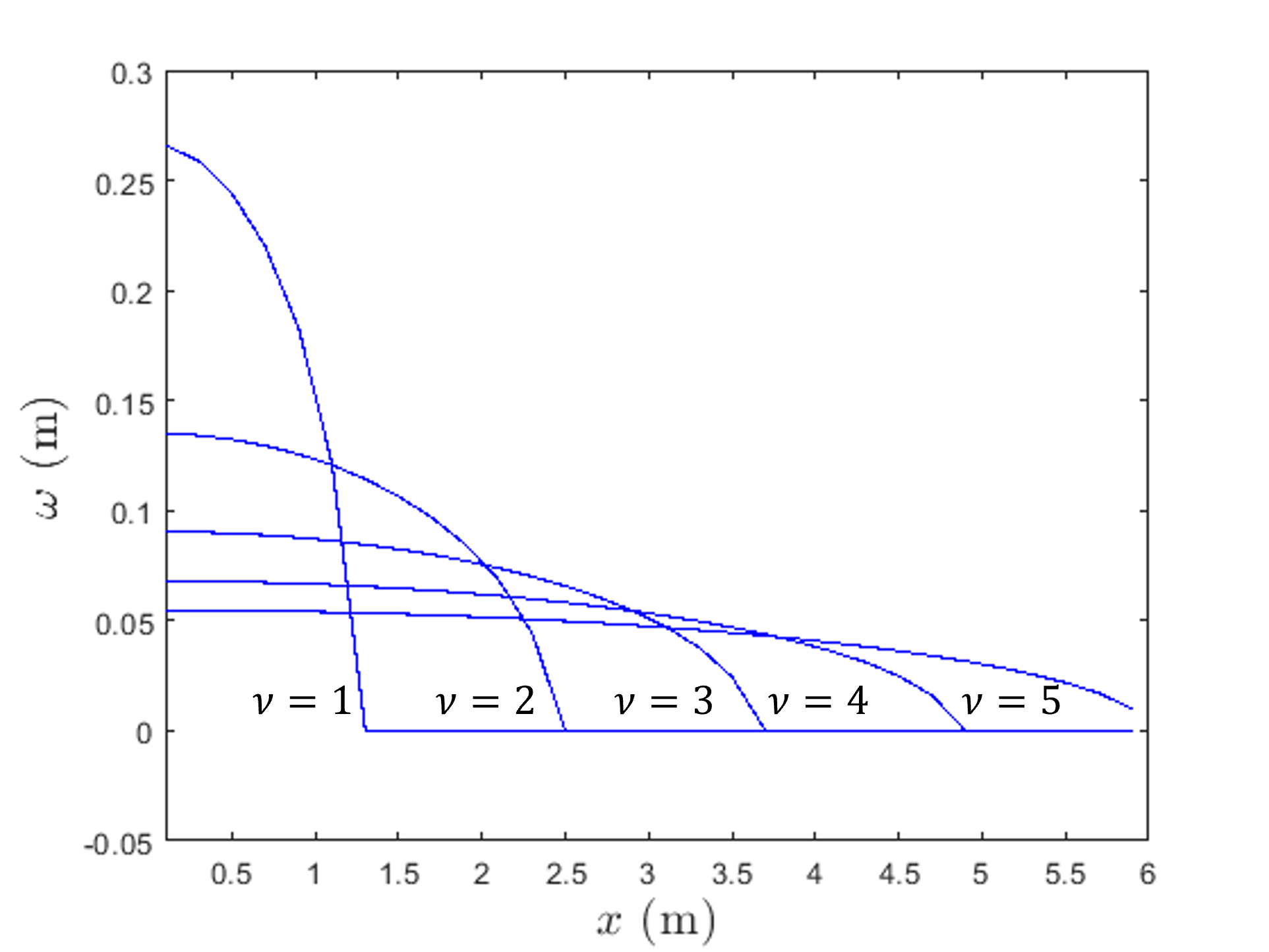}
		\caption{}
		\label{fig:quasi_newton_aperture_iter}
	\end{subfigure}
	\begin{subfigure}[t]{0.49\textwidth}
		\centering
		\includegraphics[width=1\linewidth]{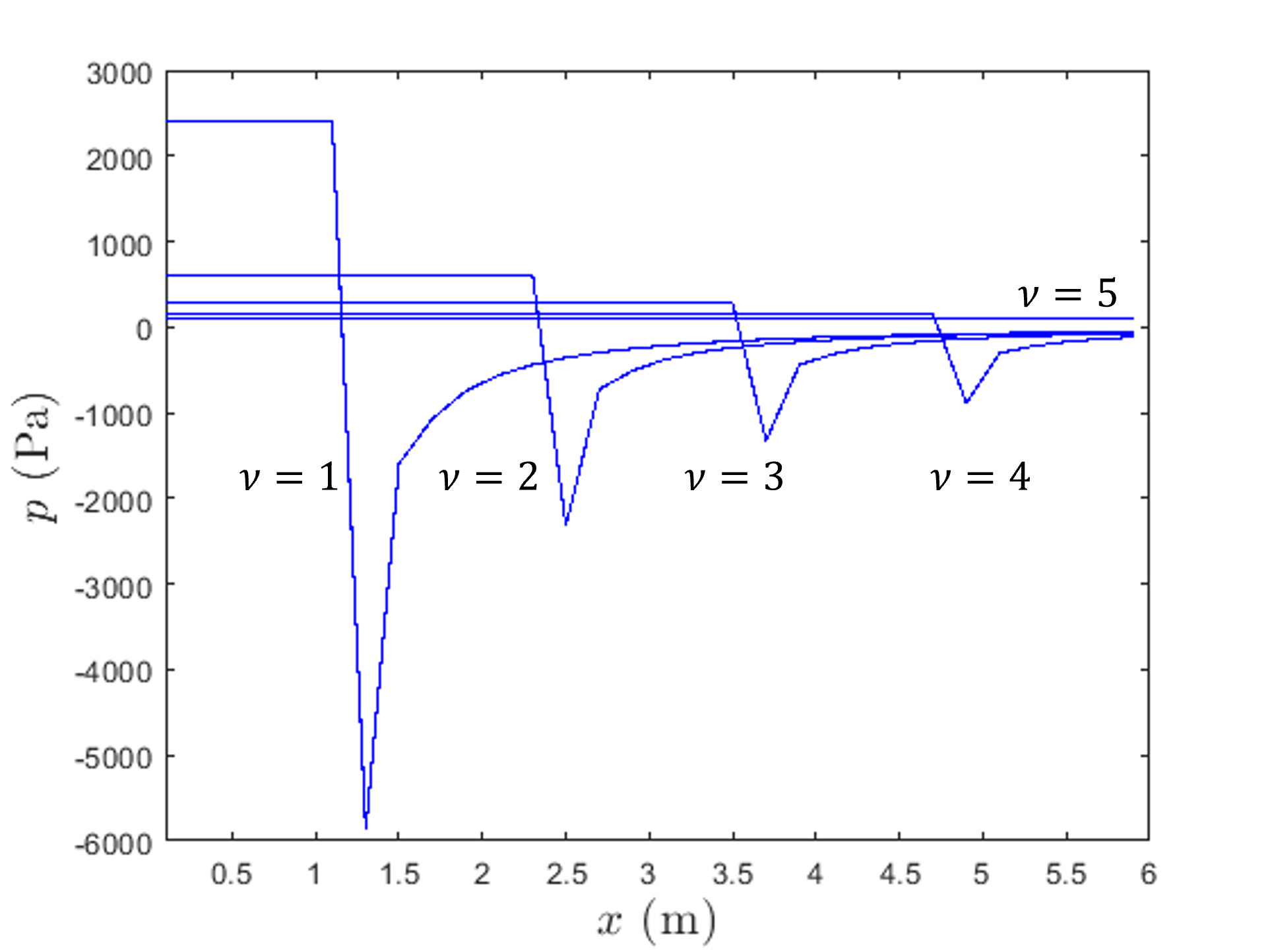}
		\caption{}
		\label{fig:quasi_newton_pressure_iter}
	\end{subfigure}
	\caption{$\omega$ and $p$ profiles during Quasi-Newton's iteration. Iteration number, $v$, increases from left to right.}
	\label{fig:quasi_newton_iter}
\end{figure}

The same simulation cases used the analysis of fixed-point stability are used except with $n_c = 15$. The results appear in \cref{fig:contraction}. In \cref{fig:lipschitz}, the Lipschitz constant $c$ is below $1$ for all simulation cases. The minimum dimensionless aperture $\frac{\omega}{\sqrt{qt}}$ is recorded during the nonlinear iteration. The dark blue area in \cref{fig:omega} indicates that all iterations produce an $\omega > 0$. In the yellow and orange regions, $\varepsilon_{0}$ is on the order of $-1e-5$, which is approximately three orders of magnitude less than the positive values of $\frac{\omega}{\sqrt{qt}}$. In summary, on the designed iteration path,  $G(\boldsymbol \omega)$ produces aperture $\omega \in D$. \cref{fig:niter} shows that the maximum number of iterations needed is below 20.
\begin{remark}
Contraction mapping is verified empirically under uniform grid size. In order to maintain the contraction in non-uniform grid systems, a constraint on the time-step size may be enforced.
 \end{remark}
\paragraph{Physical interpretations}
A simulation case using one of the dimensional groups $(\Pi_1, \Pi_2)$ in the contraction-mapping test is used to motivate the evolution of $\omega$ and $p$ during the proposed Quasi-Newton iteration. In \cref{fig:quasi_newton_iter}, $\omega$ and $p$ profiles are captured at different nonlinear iteration numbers, $\nu$. A moving fluid front in the fracture is captured for each iteration. Aperture profiles are split over two sub-regions: A fluid-filled region where $\omega > 0$ and a fluid-void region where $\omega = 0$. As the solver iterates, $\omega$ is gradually opened by the fluid front. Similarly, on the pressure profile, the fluid-filled region is represented by positive pressure, whereas the fluid-void space is indicated by negative pressure.

In terms of the structure of $\boldsymbol F$, the number of nonzero entries in matrix $\boldsymbol F$ increases as the iteration grows. For example, in the first iteration ($\nu = 1)$, only the aperture of the first grid cell is positive and, therefore, the only nonzero entries are first and second rows of $\boldsymbol F$, which means there exists flux between the first and second cells. In the second iteration, since aperture of the first two cells is positive, a third row will be added into nonzero entries of $\boldsymbol F$. Consequently, the fluid front is moving exactly one grid cell after each iteration. As a result, the number of iterations needed for convergence approximately depends on the number of fracture cells, as well as the fluid front location at the convergence. This in turn, may be interpreted to scale with the target time-size under continued injection.

\subsection{Extension and application to model DS2}\label{sec:XFEM}
The proposed Quasi-Newton formulation is extended to the co-solution of $p$ and $u$ using a general XFEM-FVM discretization. As motivated in the previous section, the contraction mapping property may be verified computationally in terms of $\omega$ as an independent-variable. In order to translate to a similar effect for aperture maintained in the coupled XFEM-FVM system, we first derive a nonlinear mapping function similar to \cref{eq:aperture_mapping}, and multiple cases are investigated. In the first case, a group of injection tests are conducted in which the Lipschitz constant $c$ is evaluated and analyzed for contraction in aperture. In the second case, the viscosity--dominated KGD problem is studied to show the robustness of the solver. In the final case, multiple fracture propagation is considered and solver performance is reported.

In the XFEM-FVM scheme, XFEM offers a relationship between $\boldsymbol \omega$ and $\boldsymbol u$ as,
\begin{equation}\label{eq:xfem_aperture}
\boldsymbol \omega = \boldsymbol B \boldsymbol{u}
\end{equation}
where $\boldsymbol B$ is a rank-deficient linear matrix. Substituting \cref{eq:xfem_aperture} into \cref{eq:residual,eq:Jacobian} yields the discretized system as well as Jacobian matrix using Quasi-Newton method. A similar form, such as \cref{quasinewton}, can be written here for the XFEM--FVM discretized system,
 \begin{equation}\label{eq:quasi_newton_xfem_edfm}
    \begin{bmatrix}
 \tilde{\boldsymbol{J}}_{ff}&  \tilde{\boldsymbol{J}}_{fm}\\ 
\tilde{\boldsymbol{J}}_{mf}& \tilde{\boldsymbol{J}}_{mm}
\end{bmatrix}^v\begin{bmatrix}
\boldsymbol p\\ 
\boldsymbol u
\end{bmatrix}^{v+1} = \begin{bmatrix}
\boldsymbol F_f\\ 
\boldsymbol F_m
\end{bmatrix}
\end{equation}
where $f, m$ indicates flow and mechanics, respectively. During iterations, $ \tilde{\boldsymbol{J}}_{ff}$ is a function of $\boldsymbol{\omega}^v$ whose values are taken from the previous iteration using \cref{eq:xfem_aperture}. In the Quasi-Newton approach, $ \tilde{\boldsymbol{J}}_{fm}$ is the same as $\boldsymbol B$ defined in \cref{eq:xfem_aperture}. The rest of the terms in \cref{eq:quasi_newton_xfem_edfm} are constant and do not change across iterations. $\boldsymbol F_f = \boldsymbol q + \boldsymbol{\omega}^n $ contains the source or sink terms while $\boldsymbol F_m$ contains the Neumann boundary information.

In order to derive a nonlinear mapping for  $\boldsymbol \omega$ in the current system, the Schur complement is constructed to demonstrate that the solution of $\omega$ by \cref{eq:quasi_newton_xfem_edfm} is equivalent to $\boldsymbol \omega$ solution obtained by using the following mapping $G$,
\begin{equation}\label{eq:apertureMapping_xfem_edfm}
    \boldsymbol{\omega} = G(\boldsymbol \omega) =  \boldsymbol{B} \tilde{\boldsymbol{J}}^{-1}_{mm}\left[ \boldsymbol F_{m} - \tilde{\boldsymbol{J}}_{mf} \boldsymbol S^{-1} (\boldsymbol F_f - \tilde{\boldsymbol{J}}_{fm} \tilde{\boldsymbol{J}}^{-1}_{mm}\boldsymbol F_m)\right]
\end{equation}
where all of terms in \cref{eq:apertureMapping_xfem_edfm} are constant except that the Schur complement $\boldsymbol S$ is a function of $\boldsymbol \omega$, which is defined as 
\begin{equation}\label{eq:schur}
    \boldsymbol S = \tilde{\boldsymbol{J}}_{ff} - \tilde{\boldsymbol{J}}_{fm}\tilde{\boldsymbol{J}}^{-1}_{mm}\tilde{\boldsymbol{J}}_{mf}
\end{equation}
Starting from \cref{eq:quasi_newton_xfem_edfm}, the deformation solution,
\begin{equation}\label{eq:sol_u}
    \boldsymbol u = \tilde{\boldsymbol{J}}^{-1}_{mm}(\boldsymbol F_m - \tilde{\boldsymbol{J}}_{mf}\boldsymbol p)
\end{equation}
and the pore-pressure solution is,
\begin{equation}\label{eq:sol_p}
    \boldsymbol p = \boldsymbol S^{-1}(\boldsymbol F_f - \tilde{\boldsymbol{J}}_{fm}\tilde{\boldsymbol{J}}^{-1}_{mm}\boldsymbol F_m)
\end{equation}
Note that substituting \cref{eq:sol_p} into \cref{eq:sol_u} and left multiplication of the RHS by $\boldsymbol B$ results in \cref{eq:apertureMapping_xfem_edfm}.
\subsubsection{Injection test}
A schematic of the case is illustrated in \cref{fig:singlefrac}. The length of the domain is $100 \text{ m}$ by $100 \text{ m}$. The fracture half-length is taken as $40 \text{ m}$. The displacement at the mid point of four edges is fixed and stresses on the boundaries are assumed to be zero. Initially, the fracture is void of fluid, i.e., $\omega^0 = 0 \text{ m}$. An injection rate of $0.001 \text{ m}^2/\text{s}$ is applied and a time step $\Delta t$ is set at $85 \text{ s}$. The Young's modulus and Poisson's ratio for the rock are $8.3 \text{ GPa}$ and $0.25$, respectively. Plane strain conditions are assumed.

Fluid is injected at the mid-point of the fracture for a single-time step $\Delta t$. The mesh sizes and fluid viscosity are varied to verify the designed iteration path offers a contraction mapping using XFEM--FVM, where $c$ with respect to $\omega$ is calculated, and the result is shown in \cref{table:xfem_fvm_contraction}. First, $c$ does not exceed $1$ in any test case. Secondly, as $\mu$ is increased or mesh size is decreased, $c$ decreases. These observations agree with the results in \cref{fig:lipschitz}.

\begin{figure}[!htb]
	\centering
	\begin{subfigure}[t]{0.6\textwidth}
		\centering
		\includegraphics[width=1\linewidth]{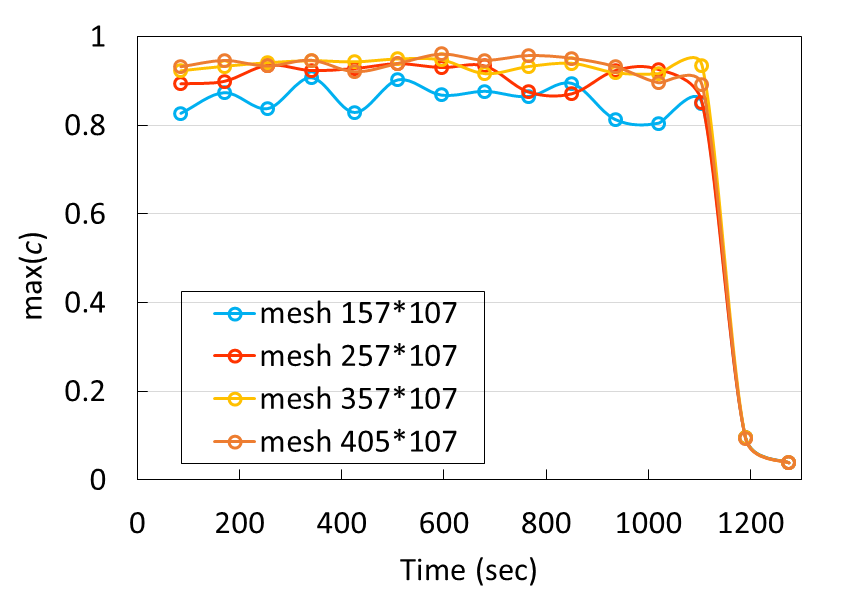}
		\caption{}
		\label{fig:injection_test_C}
	\end{subfigure}
	\begin{subfigure}[t]{0.6\textwidth}
		\centering
		\includegraphics[width=1\linewidth]{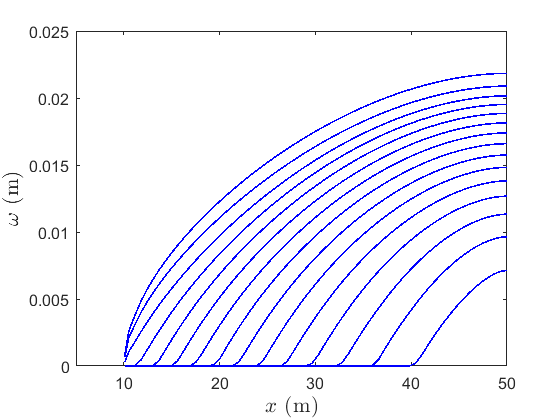}
		\caption{}
		\label{fig:injection_test_omega}
	\end{subfigure}
	\caption{(a) $\text{max}(c)$ of the case $\mu  = 20 \text{ Pa} \cdot \text{s}$ recorded at each time step solve; (b) snapshots of $\omega$ profiles over half fracture length at different times. Time increases from right to left.}
	\label{fig:injection_test}
\end{figure}

Next, time is marched forward with a fixed $\Delta t = 85 \text{ sec}$ until the injected fluid  fills the entire fracture. Four different mesh sizes in \cref{table:xfem_fvm_contraction} with $\mu = 20 \text{ Pa} \cdot \text{s}$ are evaluated. The maximum $c$ in each time step solution and snapshots of $\omega$ profiles are shown in \cref{fig:injection_test}. In \cref{fig:injection_test_omega}, during injection, $\omega$ is gradually opened  while the fluid front is moving from the injection point to the fracture tip. Before the injected fluid fills the fracture, $c$ remains below 1 under the four-tested meshes presented in \cref{fig:injection_test_C}. A higher value of $c$ is a result of subsequently opened fracture segments with each nonlinear iteration. Once the fluid front reaches the fracture tip, $c$ drops significantly below $0.1$. A significant drop of $c$ indicates an improvement of local convergence speed.

\begin{table}[htb]\centering
\caption{Maximum $c$ during nonlinear iterations (first number in the bracket) and the number of nonlinear iterations (second number in the bracket)}
  \noindent\makebox[\textwidth]{%
    \begin{tabular}{|l|*{4}{c|}}
      \hline
      \diagbox[width=\dimexpr \textwidth/5+2\tabcolsep\relax, height=1cm]{ mesh sizes }{$\mu (\text{Pa} \cdot \text{s})$  }
                   & 20 & 200 & 2000 & 20000 \\
      \hline
      $157 \times 107$ & $(0.83, 34)$& $(0.81, 25)$ & $(0.69, 25)$&  $(0.53, 17)$\\
      \hline
      $257 \times 107$ & $(0.89, 44)$ &$(0.86, 39)$ & $(0.82, 26)$& $(0.75, 26)$ \\
      \hline
      $357 \times 107$ &  $(0.92, 51)$ & $(0.89, 40)$& $(0.87, 42)$& $(0.76,35)$\\
      \hline
      $405 \times 107$ & $(0.95, 65)$ & $(0.91, 59)$& $(0.88, 37)$ &  $(0.83,41)$\\
      \hline
    \end{tabular}
  }%
  \label{table:xfem_fvm_contraction}
\end{table}
\begin{figure}[htb]
	\centering
	\includegraphics[width=0.6\linewidth]{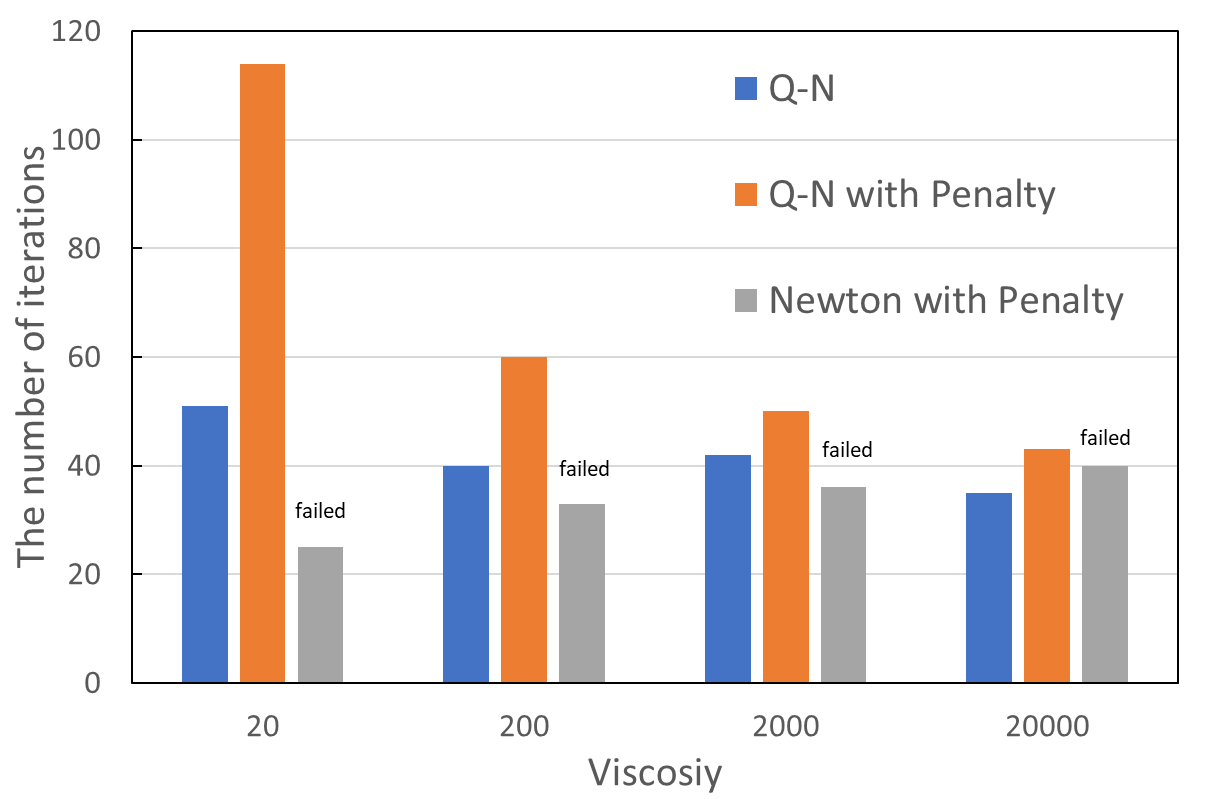}
	\caption{Comparison of three different nonlinear solvers. "failed" above the "Newton with Penalty" bar means the solver fails after a certain number of iterations. }
	\label{fig:contact}
\end{figure}
In computational mechanics, the negative aperture is addressed by applying  supporting forces on fracture surfaces, also known as contact force. The  penalty method or Lagrangian multipliers are widely-used numerical techniques. We briefly describe how the contact force is added into the coupled system in  \cref{Discretization}. However, a reckless application of contact constraints to hydro-mechanics system may not overcome the non-physical solutions.   We test  various viscosity scenarios under the mesh size $357\times 107$. Two numerical schemes are considered here: Newton with the penalty method and Quasi-Newton with the penalty method. The results show in \cref{fig:contact}. First of all, Newton's solver with the penalty method fails for all of test cases here. Newton's method yields  giant updates which destroy the nonlinear solver system. Since both physical and nonphysical solutions are stable for Newton's method, direct application of the penalty method will not guide Newton's path to the correct one and give rise to the failure of the nonlinear solver. On the other hand, all Quasi-Newton with the penalty method converges to the physical solution at a slight higher cost than the pure Quasi-Newton method. The Quasi-Newton could possibly produce very small negative aperture which will introduce the penalty method into calculation. Nevertheless, it converges to the correct solution at the end.

\subsubsection{KGD fracture propagation}
The proposed method is applied to the simulation of linear-elastic fluid-driven fracture propagation. KGD fracture is considered under two propagation regimes; viscosity-dominated and toughness-dominated. These are are controlled by the dimensionless parameter $\mathcal{K}_m$ (\cite{zeng2018fully})
\begin{equation}
    \mathcal{K}_m = \frac{8K_c}{(2E^{'3}\mu ^{'}q)^{1/4}},
\end{equation}
where $K_c$ is the critical stress intensity factor (SIF); $\mu^{'} = 12\mu$, $E'=E/(1-\nu^{2})$ is the equivalent Young's modulus in the plain strain condition; and $q$ is the flow rate into two wings of the fracture. If $\mathcal{K}_{m} < 1$, the flow lies in the viscosity-dominated regime where rock is very brittle ($K_{c}\rightarrow 0$) and energy dissipation is dominant in viscous flow. On the other hand, if $\mathcal{K}_{m} > 4$, the process becomes toughness-dominated regime where energy is mostly used to break the rock and factors from either small aperture or highly-viscous fluid could be neglected. 

\begin{figure}[!htb]
	\centering
	\begin{subfigure}[t]{0.49\textwidth}
		\centering
		\includegraphics[width=0.6\linewidth]{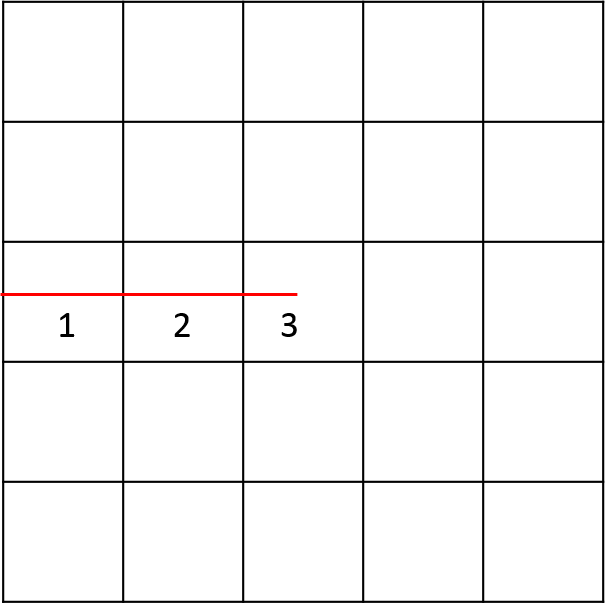}
		\caption{}
		\label{fig:before_prop}
	\end{subfigure}
	\begin{subfigure}[t]{0.49\textwidth}
		\centering
		\includegraphics[width=0.6\linewidth]{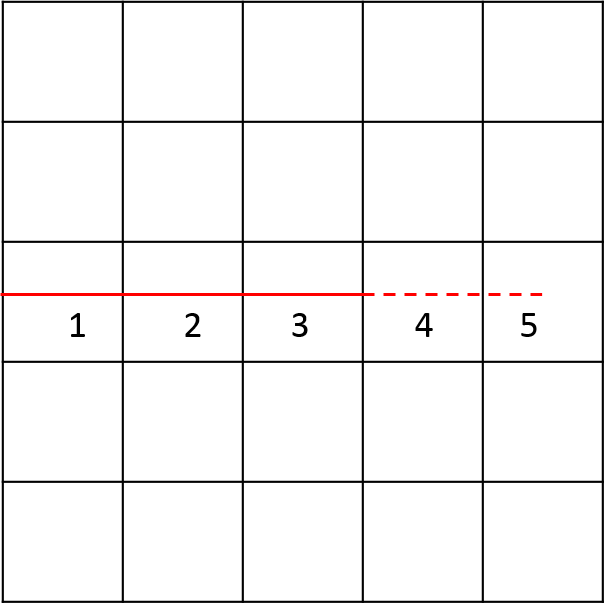}
		\caption{}
		\label{fig:after_prop}
	\end{subfigure}
	\caption{fracture grids and the numbering; (a) before propagation (b) after propagation.}
	\label{fig:prop}
\end{figure}

The rock-failure criterion is based on Irwin's law of linear elastic fracture mechanics and SIF, $K_{eq}$, is used as an indicator. $K_{eq}$ is never allowed to be larger than the critical $K_c$ during fracture propagation. When $K_{eq} < K_{c}$, fracture is static. Otherwise, fracture is extended by a certain length $\Delta a$ defined by the user. 
In terms of fracture-mesh updates, multiple fracture segments within the same background mesh grid are prohibited. In this way, multiple small fracture segments will be eliminated and large grid-size contrast in the numerical calculation can be avoided. For example, in \cref{fig:prop}, the fracture segment 3 that partially cuts the grid (\cref{fig:before_prop}) will extend to reach the boundary of the grid (\cref{fig:after_prop}), when fracture propagation is triggered. More details of the fracture propagation algorithm could be referred to  \cref{appendix3}.   

The results suggest that Newton's method is more likely to converge to the nonphysical solution when fracture propagation lies in the viscosity-dominated regime. To further investigate this, consider application of the Quasi-Newton method using the KGD analytical solution in the viscosity-dominated regime.

A square domain of dimensions $100 \times 100 \text{ m}^2$ is modeled with a Cartesian mesh of size $317 \times 117$. The initial fracture position is at the center of the domain with a half length of $2 \text{ m}$. The rock and fluid parameters are listed in \cref{table:KGDModelPropertyImplicit} and the maximum time-step size is $0.5 \text{ sec}$. The total simulation time is $90\text{ sec}$. Two advancement lengths, $\Delta a = 2, 4 \text{ m}$, are applied. The convergence criterion is controlled as,
\begin{equation}
    \sqrt{\frac{\sum_{i =1 }^{N} (\omega^{v+1} - \omega^{v})^2}{N}} < 1e-8
\end{equation}
where $N$ is the number of fracture grids.

\begin{table}[ht]
    \centering
	\caption{Input parameters for KGD model verification }
	\label{table:KGDModelPropertyImplicit}
	\begin{tabular}{cccccccc}
		\hline
		 $E$(GPa) & $\nu$ & $K_{c}(\text{MPa}\sqrt{\text{m}})$ & $\mu (\text{Pa}\cdot \text{s})$ & $q(\text{m}^{2}/\text{s})$ & $\mathcal{K}_{m}$  \\ \hline
	  8.3        & 0.25           & 0.5           &  2e-3                & 0.001                 & 0.78                \\ \hline
	\end{tabular}
\end{table}

\begin{figure}[!htb]
	\centering
	\begin{subfigure}[t]{0.45\textwidth}
		\centering
		\includegraphics[width=1\linewidth]{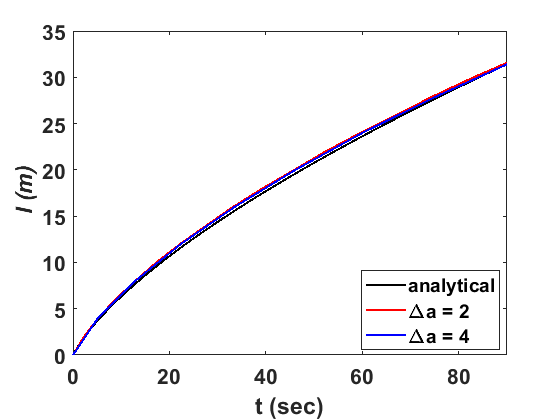}
		\caption{}
		\label{fig:length}
	\end{subfigure}
	\begin{subfigure}[t]{0.54\textwidth}
		\centering
		\includegraphics[width=1\linewidth]{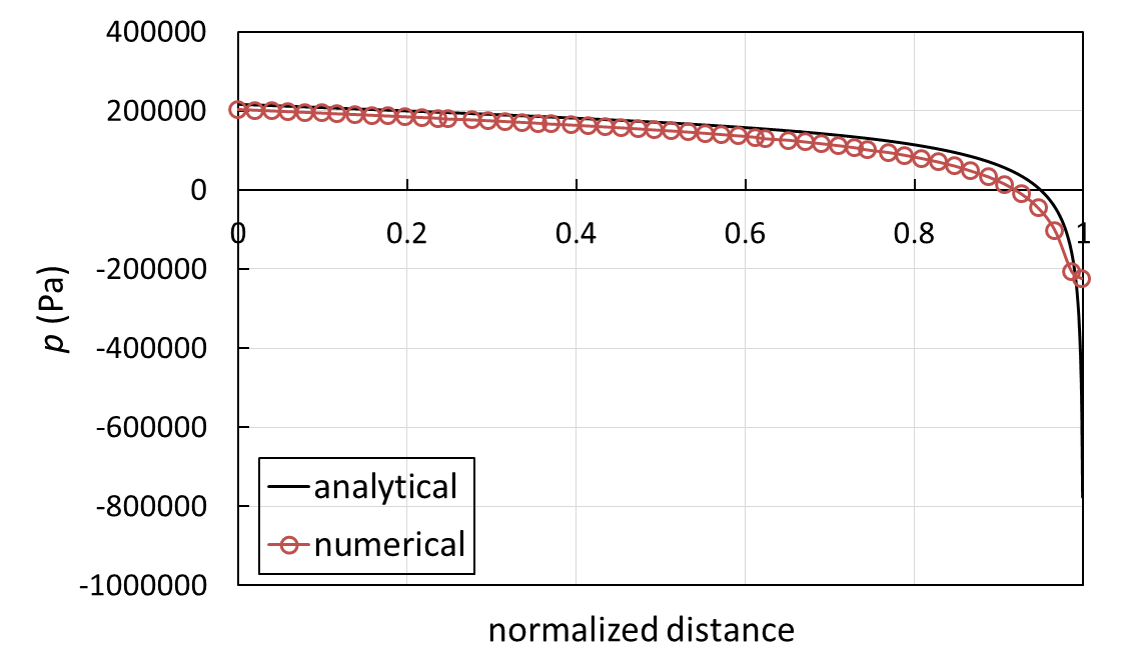}
		\caption{}
		\label{fig:pressure}
	\end{subfigure}
	\caption{(a) fracture half length evolution with time (b) $p$ profile when half fracture length is equal to $16$ m. }
	\label{fig:verification}
\end{figure}
The viscosity-dominated analytical solution is generated using the form given in \cite{adachi2002fluid}. A comparison of numerical and analytical results are illustrated in \cref{fig:verification} for both time-step sizes considered. The pressure profile of viscosity-dominated fracture propagation exhibits negative values near the tip and asymptotically decays. The singular behavior of pressure at the tip adds additional difficulty to Newton's method in obtaining this physical solution.

\begin{figure}[!htb]
	\centering
	\begin{subfigure}[t]{0.49\textwidth}
		\centering
		\includegraphics[width=1\linewidth]{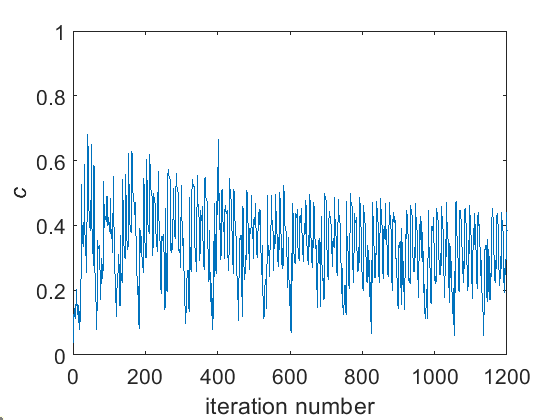}
		\caption{}
		\label{fig:a_2_c}
	\end{subfigure}
	\begin{subfigure}[t]{0.49\textwidth}
		\centering
		\includegraphics[width=1\linewidth]{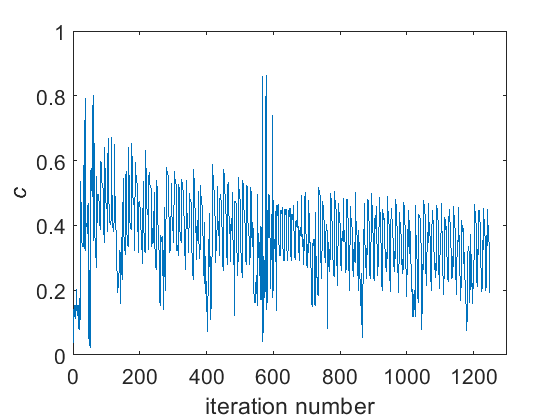}
		\caption{}
		\label{fig:a_4_c}
	\end{subfigure}
	\begin{subfigure}[t]{0.49\textwidth}
		\centering
		\includegraphics[width=1\linewidth]{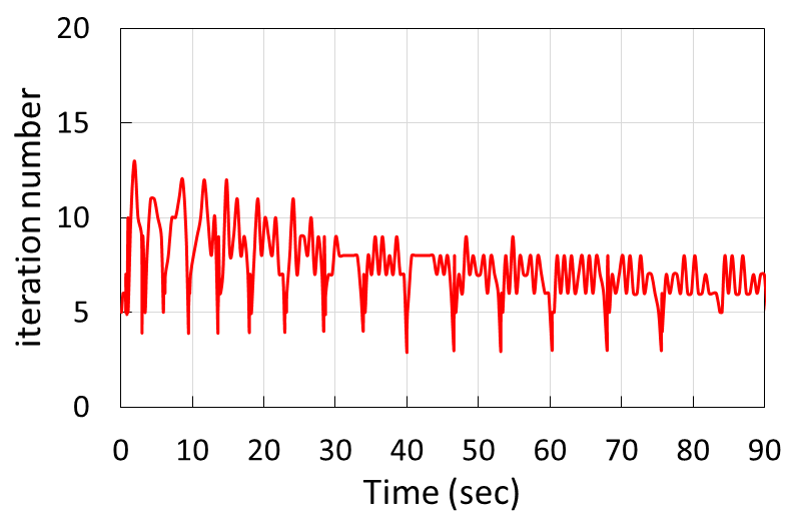}
		\caption{}
		\label{fig:a_2_iter}
	\end{subfigure}
	\begin{subfigure}[t]{0.49\textwidth}
		\centering
		\includegraphics[width=1\linewidth]{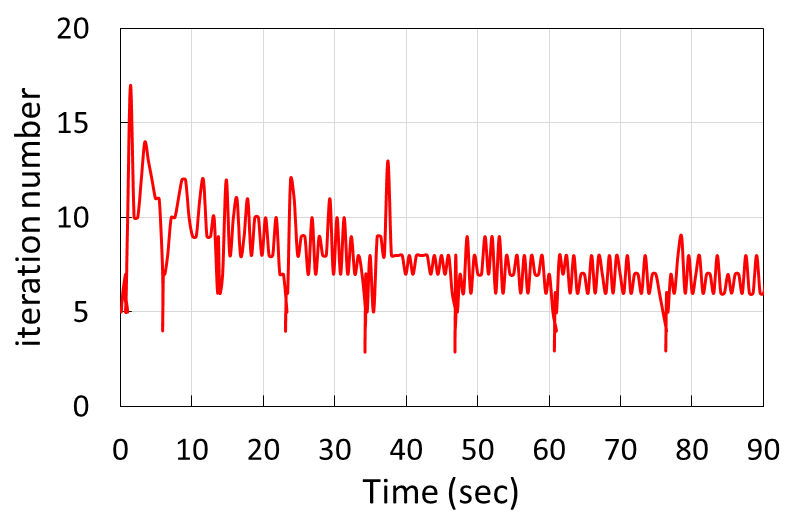}
		\caption{}
		\label{fig:a_4_iter}
	\end{subfigure}
	\caption{(a) and (b) $c$ over the simulation history (c) and (d) the number of newton iterations in each time step solve. First column corresponds to results with $\Delta a = 2$; Second column corresponds to results with $\Delta a = 4$.}
	\label{fig:visco_fp_c}
\end{figure}

During the simulation, the Quasi-Newton approach always converges to the physical solution and the performance is reported in \cref{fig:verification}. The constant $c$ computed in each iteration over the entire simulation for both step-length sizes are plotted in \cref{fig:a_2_c} and \cref{fig:a_4_c}. $c$ is below $1$, confirming the contraction mapping property on the designed iteration path throughout the course of entire simulation. The number of iterations for each time step is illustrated in \cref{fig:a_2_iter}  and \cref{fig:a_4_c}.
\subsubsection{multiple fracture propagation}
\begin{figure}[!htb]
	\centering
	\begin{subfigure}[t]{0.49\textwidth}
		\centering
		\includegraphics[width=1\linewidth]{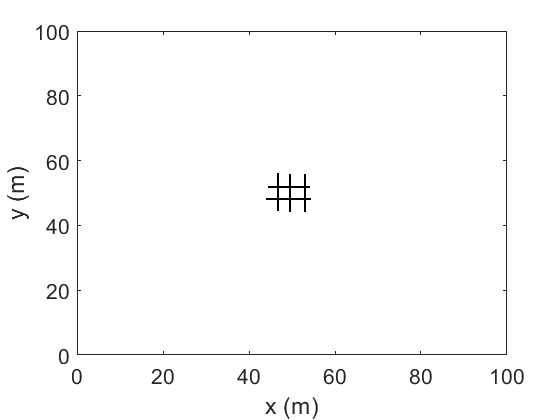}
		\caption{}
		\label{fig:mf_fp0}
	\end{subfigure}
	\begin{subfigure}[t]{0.49\textwidth}
		\centering
		\includegraphics[width=1\linewidth]{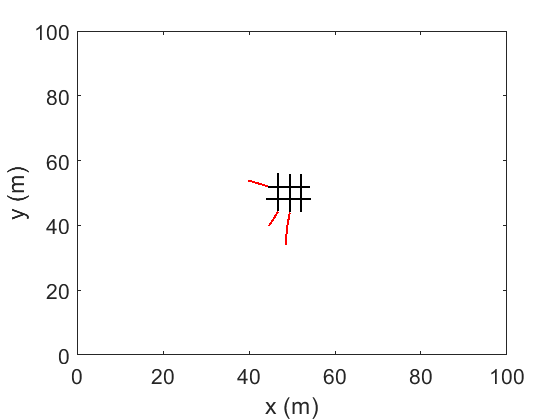}
		\caption{}
		\label{fig:mf_fp1}
	\end{subfigure}
	\begin{subfigure}[t]{0.49\textwidth}
		\centering
		\includegraphics[width=1\linewidth]{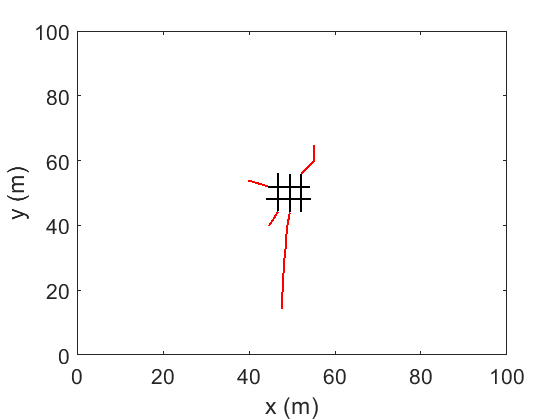}
		\caption{}
		\label{fig:mf_fp2}
	\end{subfigure}
	\begin{subfigure}[t]{0.49\textwidth}
		\centering
		\includegraphics[width=1\linewidth]{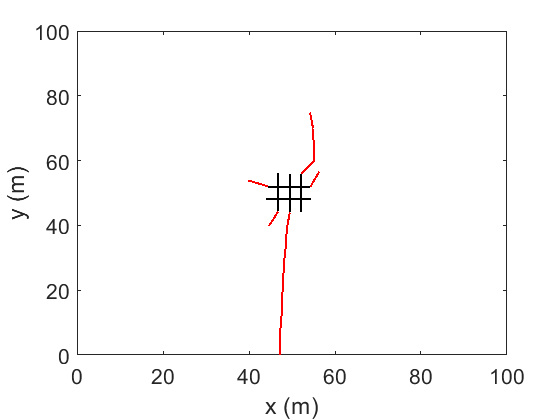}
		\caption{}
		\label{fig:mf_fp3}
	\end{subfigure}
	\caption{(a) initial fracture configuration; (b) (c) and (d)  fracture configurations at three different times in chronological order.}
	\label{fig:mf_fp}
\end{figure}

Consider a pre-existing fracture network as illustrated in \cref{fig:mf_fp0}. In this test case, point source fluid injection into the network and fracture propagation are simulated under various conditions. The input parameters are listed in \cref{table:ipmf}. The upper and lower boundaries are under the maximum horizontal stress $\sigma_H$ while the left and right boundaries are under the minimum one $\sigma_h$. The point sources are located on the two horizontal fractures at $(50,48)$ and $(50, 52)$, respectively. The injection rates that equal $0.001 \text{ m}^{2}/\text{s}$ are the same for two point sources. At $t = 0$, fractures are void of fluid. The simulation time lasts $122$ secs.
\begin{table}[!htb]
\centering
\caption{Input parameters for the case of multiple fracture propagation}
\label{table:ipmf}
\begin{tabular}{lll}
\hline
 & value& unit\\ \hline
 $E$  & $8.3$ & GPa\\ \hline
$\nu$ & 0.25 &\\ \hline
$\mu$    & 2e-3 & $\text{Pa}\cdot \text{s}$\\ \hline
 $K_{c}$   & 2e6 & $\text{Pa} \sqrt{\text{m}}$ \\ \hline
 $\Delta a$  & 5& m \\ \hline
 $\sigma_H$ & 0.4 & MPa \\  \hline
 $\sigma_h$ & 0.2 & MPa \\ \hline
\end{tabular}
\end{table}

\begin{figure}[!htb]
	\centering
	\includegraphics[width=0.7\linewidth]{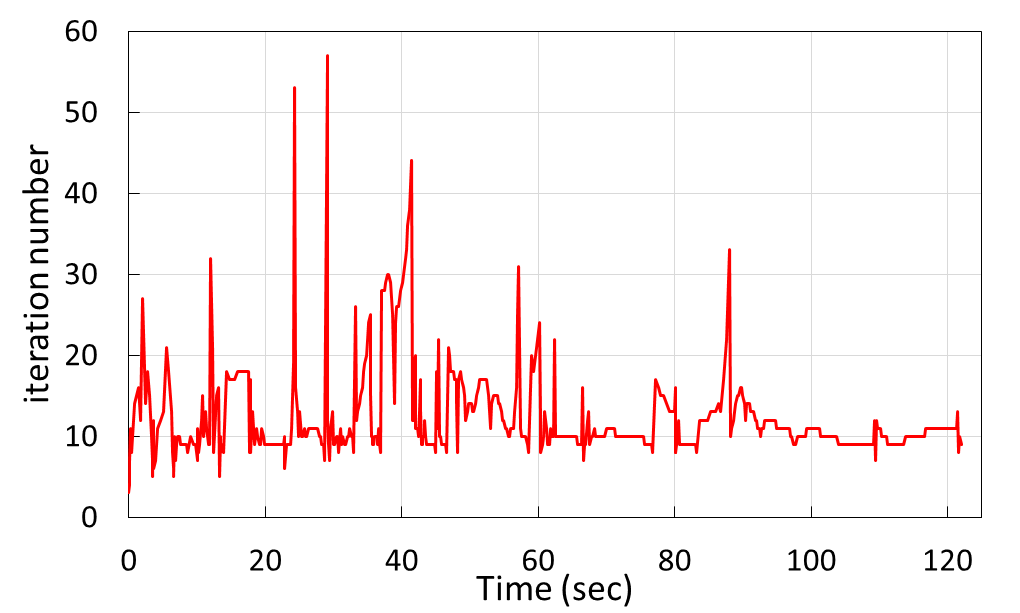}
	\caption{nonlinear iterations for each time step solve}
	\label{fig:mf_niter}
\end{figure}

\begin{figure}[!htb]
	\centering
	\includegraphics[width=0.55\linewidth]{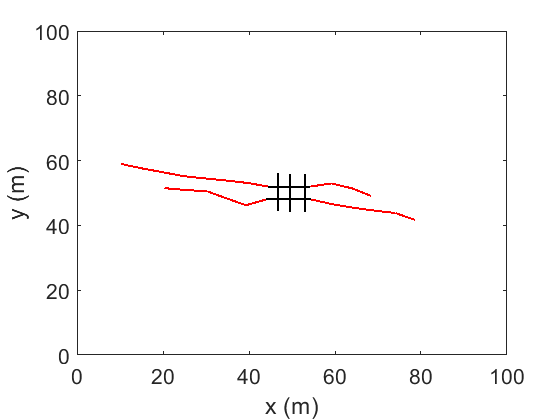}
	\caption{fracture geometry when $\sigma_{H}$ is in the horizontal direction and $\sigma_{h}$ is in the vertical direction.}
	\label{fig:rotate_stress_bc}
\end{figure}
The plots in \cref{fig:mf_fp1,fig:mf_fp2,fig:mf_fp3} capture three snapshots during fracture propagation. At the earliest time, injected fluid first opens two horizontal fractures while the other three vertical fractures are closed due to stresses on the boundaries. In \cref{fig:mf_fp1,fig:mf_fp2}, as three vertical fractures have the least resistance from far field stresses than the two horizontal ones, vertical fractures are gradually opened by fluid and start to propagate in the vertical direction. In the meantime, only one horizontal fracture starts to propagate.  In the latest time shown in \cref{fig:mf_fp3}, only propagation of the vertical fractures is observed in the vertical direction perpendicular to the minimum principal stress. The number of nonlinear iterations is shown in \cref{fig:mf_niter}. There are a few times when the iteration number is above 30 since  nonuniform-spaced fracture grids are inevitably introduced during fracture propagation . On the other hand, Newton's method is also applied to this case and fails in the first time when a fracture starts to propagate.

We also rotate  $\sigma_{H}$ and $\sigma_{h}$ on the boundaries and run the same simulation case. In this regard, only two horizontal fractures are advancing in the x direction while three vertical fractures remain static in \cref{fig:rotate_stress_bc}.

\section{Discussion and Conclusion}\label{sec:conclusion}
A Quasi-Newton approach is proposed to avoid convergence to nonphysical solutions and to improve solver robustness for coupled hydro-mechanics and fracture propagation simulation. Supported by thorough empirical observation and some analysis, the following findings are listed:
\begin{itemize}
    \item multiple solutions may exist for coupled hydro-mechanical models of Poiseiulle flow in fracture, and this has been observed in fully discrete and semi-analytical models.
    \item the derivative of the flux function with respect to aperture in the Jacobian matrix can influence the Newton path towards nonphysical solution. 
    \item empirical stability analysis suggests that the physical solution is the only stable fixed point for the proposed Quasi-Newton method, whereas both physical and nonphysical solutions are stable fixed points for Newton's method.
    \item the proposed Quasi-Newton method is a contraction for aperture on uniform grids.
\end{itemize}

In fracture propagation, the simulation meshing is adjusted with simulation time. An inappropriate initialization of unknowns in the newly-formed fracture space can possibly steer the nonlinear Newton solution process for state (displacement and pressure) to a nonphysical fixed-point attractor. This more frequently observed in viscosity-dominated propagation problems. The proposed Quasi-Newton method is demonstrated to overcome these issues in viscosity-dominated KGD fracture propagation, as well as under the simultaneous propagation of multiple fractures. The proposed method has been tested under the setting of a 2D domain, but can be readily extended to 3D domain.
\section{Acknowledgements}
This material is based upon work supported by the U.S. Department of Energy under Award Number DE-FE-0031777. The authors also acknowledge partial funding from the members of the  TU Future Reservoir Simulation Systems \& Technology (FuRSST) Industry-University Consortium.

\bibliographystyle{plain}
\bibliography{mybibfile}

\begin{thebibliography}{10}

\bibitem{adachi2002fluid}
Jos{\'e}~Ignacio Adachi.
\newblock Fluid-driven fracture in permeable rock.
\newblock {\em PhD thesis}, 2002.

\bibitem{bavzant2014fracking}
Zden{\v{e}}k~P Ba{\v{z}}ant, Marco Salviato, Viet~T Chau, Hari Viswanathan, and
  Aleksander Zubelewicz.
\newblock Why fracking works.
\newblock {\em Journal of Applied Mechanics}, 81(10), 2014.

\bibitem{breede2013systematic}
Katrin Breede, Khatia Dzebisashvili, Xiaolei Liu, and Gioia Falcone.
\newblock A systematic review of enhanced (or engineered) geothermal systems:
  past, present and future.
\newblock {\em Geothermal Energy}, 1(1):1--27, 2013.

\bibitem{campos2017stability}
Beatriz Campos, Alicia Cordero, Juan~R Torregrosa, and Pura Vindel.
\newblock Stability of king’s family of iterative methods with memory.
\newblock {\em Journal of Computational and Applied Mathematics}, 318:504--514,
  2017.

\bibitem{edwards2015model}
Ryan~WJ Edwards, Michael~A Celia, Karl~W Bandilla, Florian Doster, and
  Cynthia~M Kanno.
\newblock A model to estimate carbon dioxide injectivity and storage capacity
  for geological sequestration in shale gas wells.
\newblock {\em Environmental science \& technology}, 49(15):9222--9229, 2015.

\bibitem{girault2016convergence}
Vivette Girault, Kundan Kumar, and Mary~F Wheeler.
\newblock Convergence of iterative coupling of geomechanics with flow in a
  fractured poroelastic medium.
\newblock {\em Computational Geosciences}, 20(5):997--1011, 2016.

\bibitem{girault2015lubrication}
Vivette Girault, Mary~F Wheeler, Benjamin Ganis, and Mark~E Mear.
\newblock A lubrication fracture model in a poro-elastic medium.
\newblock {\em Mathematical Models and Methods in Applied Sciences},
  25(04):587--645, 2015.

\bibitem{gordeliy2013implicit}
Elizaveta Gordeliy and Anthony Peirce.
\newblock Implicit level set schemes for modeling hydraulic fractures using the
  xfem.
\newblock {\em Computer Methods in Applied Mechanics and Engineering},
  266:125--143, 2013.

\bibitem{gupta2018coupled}
P~Gupta and Carlos~Armando Duarte.
\newblock Coupled hydromechanical-fracture simulations of nonplanar
  three-dimensional hydraulic fracture propagation.
\newblock {\em International Journal for Numerical and Analytical Methods in
  Geomechanics}, 42(1):143--180, 2018.

\bibitem{hunsweck2013finite}
Michael~J Hunsweck, Yongxing Shen, and Adri{\'a}n~J Lew.
\newblock A finite element approach to the simulation of hydraulic fractures
  with lag.
\newblock {\em International Journal for Numerical and Analytical Methods in
  Geomechanics}, 37(9):993--1015, 2013.

\bibitem{lipnikov2007monotone}
Konstantin Lipnikov, Mikhail Shashkov, Daniil Svyatskiy, and Yu~Vassilevski.
\newblock Monotone finite volume schemes for diffusion equations on
  unstructured triangular and shape-regular polygonal meshes.
\newblock {\em Journal of Computational Physics}, 227(1):492--512, 2007.

\bibitem{lipnikov2009interpolation}
Konstantin Lipnikov, Daniil Svyatskiy, and Yuri Vassilevski.
\newblock Interpolation-free monotone finite volume method for diffusion
  equations on polygonal meshes.
\newblock {\em Journal of Computational Physics}, 228(3):703--716, 2009.

\bibitem{liu2020modeling}
Fushen Liu.
\newblock Modeling hydraulic fracture propagation in permeable media with an
  embedded strong discontinuity approach.
\newblock {\em International Journal for Numerical and Analytical Methods in
  Geomechanics}, 44(12):1634--1655, 2020.

\bibitem{liu2019history}
Zhe Liu and Albert~C Reynolds.
\newblock History matching an unconventional reservoir with a complex fracture
  network.
\newblock In {\em SPE Reservoir Simulation Conference}. OnePetro, 2019.

\bibitem{ren2021integrated}
Guotong Ren and Rami~M Younis.
\newblock An integrated numerical model for coupled poro-hydro-mechanics and
  fracture propagation using embedded meshes.
\newblock {\em Computer Methods in Applied Mechanics and Engineering},
  376:113606, 2021.

\bibitem{spence1985self}
DA~Spence and P~Sharp.
\newblock Self-similar solutions for elastohydrodynamic cavity flow.
\newblock {\em Proceedings of the Royal Society of London. A. Mathematical and
  Physical Sciences}, 400(1819):289--313, 1985.

\bibitem{terekhov2017cell}
Kirill~M Terekhov, Bradley~T Mallison, and Hamdi~A Tchelepi.
\newblock Cell-centered nonlinear finite-volume methods for the heterogeneous
  anisotropic diffusion problem.
\newblock {\em Journal of Computational Physics}, 330:245--267, 2017.

\bibitem{xu2021revisiting}
Shiqian Xu, Guotong Ren, Rami~M Younis, and Qihong Feng.
\newblock Revisiting field estimates for carbon dioxide storage in depleted
  shale gas reservoirs: The role of geomechanics.
\newblock {\em International Journal of Greenhouse Gas Control}, 105:103222,
  2021.

\bibitem{zeng2018fully}
Qinglei Zeng, Zhanli Liu, Tao Wang, Yue Gao, and Zhuo Zhuang.
\newblock Fully coupled simulation of multiple hydraulic fractures to propagate
  simultaneously from a perforated horizontal wellbore.
\newblock {\em Computational Mechanics}, 61(1):137--155, 2018.

\end{thebibliography}
\appendix
\section{}\label{appendix2}
Matrix $\boldsymbol F$ from \cref{nonlinearSystem} is 
\begin{equation}
     \boldsymbol{F} = \frac{\Delta t}{12\mu\Delta x^2} \begin{pmatrix}
\omega^3_{1 + \frac{1}{2}} &-\omega^3_{1 + \frac{1}{2}}  & 0 & 0 & \hdots  & 0\\ 
 -\omega^3_{1 + \frac{1}{2}} & \omega^3_{1 + \frac{1}{2}} + \omega^3_{2 + \frac{1}{2}} & -\omega^3_{2 + \frac{1}{2}} & 0 &\hdots & 0 \\ 
\vdots  & \vdots &  \vdots&\vdots &  \ddots & \vdots\\ 
0& 0 &0 & 0 & - \omega^3_{n-\frac{1}{2}} & \omega^3_{n-\frac{1}{2}} 
\end{pmatrix} 
\end{equation}
where each row and column sum of $\boldsymbol F$ is $0$.
\section{}\label{appendix1}
Proof of Proposition \ref{prop1}
\begin{proof}
According to \cref{quasinewton}
\begin{equation}\label{prop1_1}
 \boldsymbol{A}\boldsymbol{p}^{*} + \boldsymbol{F}(\boldsymbol{p})\boldsymbol{p}^{*}= \boldsymbol{q} + \boldsymbol{\omega}^{n}    
\end{equation}
Take the column sum of \cref{prop1_1}, we get
\begin{equation}
    \sum_i (\boldsymbol{A}\boldsymbol{p}^{*})_i + \sum_i(\boldsymbol{F}(\boldsymbol{p})\boldsymbol{p}^{*})_i =  \sum_i(\boldsymbol{q} + \boldsymbol{\omega}^{n})_i
\end{equation}
Since $\forall \boldsymbol{p} \in \mathbb{R}^{n_c}$, $\sum_i\big(\boldsymbol{F}(\boldsymbol{p})\big)_{ij} = \sum_j\big(\boldsymbol{F}(\boldsymbol{p})\big)_{ij} = 0$, then  $\sum_i(\boldsymbol{F}(\boldsymbol{p})\boldsymbol{p}^{*})_i = 0$. Suppose the linear solver can yield the exact solution of $\boldsymbol{p}^{*}$, then Proof \ref{prop1} is valid.
\end{proof}

\section{}\label{Discretization}
The discretization of the proposed the coupled FVM-XFEM will be briefly described in this section. First, flow in the fracture is handled by the finite difference method and its discretized form is the same as \cref{eq:quasi-newton}. On the other side, the displacement approximation of XFEM is
\begin{equation}
    \label{interpolation1}
\begin{aligned}
	\bm{u} &= \sum_{i\in I}N_{i}\bar{\bm{u}}_{i} + \sum_{i\in L}N_{i}(H_{c} - H^{i}_{c})\bar{\bm{a}}_{i}  + \sum_{i \in K}\sum_{l = 1}^{4}N_{i}(F_{l} - F^{i}_{l})\bar{\bm{b}}^{l}_{i}
\end{aligned}\\
\end{equation}
where $I, L, K$ is the set of standard nodes, Heaviside-enriched nodes and tip-enriched nodes, respectively. $N_i$ is the shape function, $H_c$ is the Heaviside function, and $F_l$ is the tip enrichment function. Expressions for the function above are listed in \cite{ren2021integrated}. $\bar{\bm{u}}_{i}$ is the standard nodal displacement, $\bar{\bm{a}}_{i}$ is Heaviside-enriched nodal displacement, and $\bar{\bm{b}}^{l}_{i}$ is tip-enriched nodal displacement. Starting from the weak form of \cref{GeomechanicsSF} 
\begin{equation}\label{weakform1}
\int_{\Omega}  \delta\bm{\varepsilon} : \bm{\sigma} d\bm{x} = \int_{\Gamma_{t}}\delta \bar{\bm{u}} \cdot \bm{t}d\bm{x} + \int_{\mathcal{C}}\llbracket \delta \bar{\bm{u}}\rrbracket  \cdot p_{F}\bm{n}_{c}d\bm{x}
\end{equation}
where $\delta \boldsymbol{\varepsilon}$ and $\delta \bar{\boldsymbol{u}}$ are the trial functions.
 Substitution of \cref{interpolation1} into the weak form  obtains
\begin{equation}
    (\delta \bar{\boldsymbol{u}})^T\sum_{e=1}^{N} \int_{\Omega^{e}} (\bm{B}_{i}^{r})^{\rm{T}}D\bm{B}_{j}^{s}d\bm{x}\bar{\boldsymbol{u}} =   (\delta \bar{\boldsymbol{u}})^T\sum_{e=1}^{N_c}\bigg( \int_{\mathcal{C}^e}N_{i}\llbracket F_{l}\rrbracket p_{F}\bm{n}_{c}d\bm{x} + \int_{\mathcal{C}^{e}} N_{i}\llbracket H_{\bm{\gamma}_{c}}\rrbracket p_{F} \bm{n}_{c}d\bm{x} \bigg) +   (\delta \bar{\boldsymbol{u}})^T\sum_{e =1}^{N_\gamma}\int_{\Gamma^{e}_{t}}N_{i}\bm{t}d\bm{x}
\end{equation}
where $r,s = u,a,b^{l}$, $B^{r,s}$ are the (enriched) shape functions derivatives with respect to coordinates. $N$ is the total number of grid blocks, $N_c$ is the number of fracture grids and $N_\gamma$ is the number of grids along the domain outer boundaries.

The aperture $\omega$ can be derived from \cref{interpolation1}
\begin{equation}
    	\omega = 2\sum_{i\in L}N_{i}\bar{\bm{a}}_{i}  + \sum_{i \in K}\sum_{l = 1}^{4}N_{i}(F^{+}_{l} - F^{-}_{l} )\bar{\bm{b}}^{l}_{i}
\end{equation}
 \cref{eq:quasi_newton_xfem_edfm} can be obtained using the discretized form of XFEM and FVM. The resultant Newton's linearized system reads
 \begin{equation}\label{fcsytem}
\begin{pmatrix}
J_{ff}& J_{fm} \\
J_{mf} & J_{mm} \\
\end{pmatrix}\begin{pmatrix}
\delta p\\ 
\delta \boldsymbol{u}\\
\end{pmatrix}=-\begin{pmatrix}
R_f\\ 
R_m\\ 
\end{pmatrix},
 \end{equation}
 Where $J_{ff}, J_{fm}, J_{mf}, J_{mm}$ are residuals' derivatives with respect to pressure and displacement. The quasi-newton method can be achieved by modification of $J_{fm}$.
 To add contact force into the system, we adopt the penalty method and the weak form now becomes 
\begin{equation}\label{weakform2}
\int_{\Omega}  \delta\bm{\varepsilon} : \bm{\sigma} d\bm{x} = \int_{\Gamma_{t}}\delta \bar{\bm{u}} \cdot \bm{t}d\bm{x} + \int_{\mathcal{C}}\llbracket \delta \bar{\bm{u}}\rrbracket  \cdot p_{F}\bm{n}_{c}d\bm{x} + \int_{\mathcal{C}}\llbracket \delta \bar{\bm{u}}\rrbracket  \cdot (-K_{ct}\omega) \bm{n}_{c}d\bm{x}
\end{equation}
 where $K_{ct}$ is the stiffness of the contact. Hence, the discretized form becomes
\begin{align}
    (\delta \bar{\boldsymbol{u}})^T\sum_{e=1}^{N} \int_{\Omega^{e}} (\bm{B}_{i}^{r})^{\rm{T}}D\bm{B}_{j}^{s}d\bm{x}\bar{\boldsymbol{u}} =   (\delta \bar{\boldsymbol{u}})^T\sum_{e=1}^{N_c}\bigg( \int_{\mathcal{C}^e}N_{i}\llbracket F_{l}\rrbracket p_{F}\bm{n}_{c}d\bm{x} + \int_{\mathcal{C}^{e}} N_{i}\llbracket H_{\bm{\gamma}_{c}}\rrbracket p_{F} \bm{n}_{c}d\bm{x} \bigg) +   (\delta \bar{\boldsymbol{u}})^T\sum_{e =1}^{N_\gamma}\int_{\Gamma^{e}_{t}}N_{i}\bm{t}d\bm{x} \\
    +(\delta \bar{\boldsymbol{u}})^T\sum_{e=1}^{N_{ct}}\bigg( \int_{\mathcal{C}^e}N_{i}\llbracket F_{l}\rrbracket (-K_{ct}\omega) \bm{n}_{c}d\bm{x} + \int_{\mathcal{C}^{e}} N_{i}\llbracket H_{\bm{\gamma}_{c}}\rrbracket (-K_{ct}\omega) \bm{n}_{c}d\bm{x} \bigg) 
\end{align}
where $N_{ct}$ denotes the number of fracture elements whose $\omega$ are negative.
 \section{} \label{appendix3}
The fracture propagation algorithm of \cite{ren2021integrated} will be briefly described in this section. The goal is to capture the time spot $t_c$ when the stress intensity factor reaches the critical value,
\begin{equation}\label{criterion}
     K_{eq}= K_c
\end{equation}
Two time-step adjustment mechanisms are involved here,
\begin{equation}
\Delta t^{n+1} := \left\{\begin{matrix}
 \alpha\Delta t^{n}& K_{eq} < K_{c} \\ 
\frac{K_{c}-K_{eq}^{old}}{K_{eq} - K_{eq}^{old}}\Delta t^{n} & K_{eq} > K_{c}
\end{matrix}\right.
\end{equation}
where $\alpha$ is the growth factor. $t^{n + 1} = t^{n} + \Delta t^{n}$. $K_{eq}^{old}$ and $K_{eq}$ is recorded at $t^{n}$ and $t^{n+1}$, respectively. For each time-step solve, the coupled system \cref{fcsytem} is solved till \cref{criterion} is achieved.

\end{document}